\documentclass[twocolumn,showpacs,amsmath,amssymb,floatfix,pra]
{revtex4}

\usepackage{graphicx}
\usepackage{bm}

\newcommand{\PM}{\raisebox{2.5pt}{$+$}\hspace{-2ex}%
\raisebox{-2.5pt}{{\scriptsize(}$\!-\!$\scriptsize{)}}}
\newcommand{\MP}{\raisebox{2.5pt}{$-$}\hspace{-2ex}%
\raisebox{-2.5pt}{{\scriptsize(}$\!+\!$\scriptsize{)}}}

\begin{document}

\title{Body-assisted van der Waals interaction between two atoms}

\author{Hassan Safari}
\author{Stefan Yoshi Buhmann}
\author{Dirk-Gunnar Welsch}
\affiliation{Theoretisch-Physikalisches Institut,
Friedrich-Schiller-Universit\"{a}t Jena,
Max-Wien-Platz 1, 07743 Jena, Germany}

\author{Ho Trung Dung}
\affiliation{Institute of Physics,
Academy of Sciences and
Technology, 1 Mac Dinh Chi Street, District 1, Ho Chi Minh city,
Vietnam}

\date{\today}

\begin{abstract}
Using fourth-order perturbation theory, a general formula for the van
der Waals potential of two neutral, unpolarized, ground-state atoms
in the presence of an
arbitrary arrangement of dispersing and absorbing magnetodielectric
bodies is derived. The theory is applied to two atoms in bulk
material and in front of a planar multilayer system, with special
emphasis on the cases of a perfectly reflecting plate and a
semi-infinite half space. It is demonstrated that the enhancement and
reduction of the two-atom interaction due to the presence of a
perfectly reflecting plate can be understood, at least in the
nonretarded limit, by using the method of image charges. For the
semi-infinite half space, both analytical and numerical results are
presented.
\end{abstract}


\maketitle


\section{Introduction}
\label{Sec:Intro}

The dispersive interaction between two neutral, unpolarized,
ground-state atoms---commonly known as the van der Waals (vdW)
interaction---may be regarded, in the nonretarded limit, i.e., for
small interatomic separations, as the mutual interaction of the
fluctuating electric dipole moments of the atoms in the ground state.
It was first calculated in this limit by London using perturbation
theory, the leading-order result being an attractive potential
proportional to $r^{-6}$, where $r$ denotes the interatomic separation
\cite{lon}. In the retarded limit, i.e., for large interatomic
separations, the interaction is due to the ground-state fluctuations
of both the atomic dipole moments and the electromagnetic far field.
This was first demonstrated by Casimir and Polder, who identified the
vdW interaction as the position-dependent shift of the system's
ground-state energy due to the coupling between the atoms and the
electromagnetic field \cite{c-p}. Using a normal-mode expansion of the
electromagnetic field and calculating the energy shift in
leading-order perturbation theory, they generalized the (nonretarded)
London potential to arbitrary distances between the two atoms, where
in particular in the retarded limit the potential was found to vary as
$r^{-7}$.

The theory has been extended in many respects, and various factors
affecting the vdW interaction have been studied. Based on a
calculation of photon scattering amplitudes, Feinberg and Sucher
extended the theory to magnetically polarizable atoms \cite{feinberg}.
They found that the vdW interaction of two magnetically polarizable
atoms is again attractive, while for two atoms of opposed type---one
magnetically and one electrically polarizable---a repulsive vdW force
may be observed. Later on, it was demonstrated that in the case of
two atoms of opposed type the nonretarded potential is proportional
to $r^{-4}$, in contrast to the $r^{-6}$-dependence of the nonretarded
potential of equal-type atoms \cite{Farina02}. The Feinberg-Sucher
result was extended to particles exhibiting crossed polarizabilities
\cite{eli}. Further studies have also included the cases of one
\cite{pass1} or both atoms \cite{p-t3,shr} being excited, leading to
potentials that vary as $r^{-6}$ and $r^{-2}$ in the nonretarded and
retarded limits, respectively. Thermal photons present for any nonzero
temperature have been shown to lead, in the retarded limit, to a
change of the vdW potential of two ground-state atoms from a $r^{-7}$-
to a $r^{-6}$-dependence as soon as the interatomic separation exceeds
the wavelength of the dominant photons \cite{nin,wen,gdk,bar}.
Modifications of the vdW interaction due to external fields have been
shown to lead to a potential varying as $r^{-3}$ in the nonretarded
limit when the applied field is unidirectional \cite{mil}.
Generalizations of the vdW interaction to the three-
\cite{Axilrod43,Aub60,Cirone96,pass2} and $N$-atom case
\cite{p-t1,p-t2} were addressed first in the nonretarded limit and
later for arbitrary interatomic separations, where the potentials were
seen to depend on the relative positions of the atoms in a rather
complicated way.

Van der Waals interactions play an important role in the understanding
of many phenomena---mostly in the field of surface science, such as
surface tension \cite{Ninham97,Bostroem01}, adhesion
\cite{Rabinowicz65}, and capillarity \cite{Rowlinson02}, but also in
chemical physics, such as colloidal interactions
\cite{Ninham97,Bostroem01b} and stability \cite{Russel89}. However,
application of the theoretical results to these phenomena requires
taking into account the influence of media on the atom-atom
interaction. An expression for the vdW interaction of two ground-state
atoms in the presence of dielectric media was first obtained by
Mahanty and Ninham based on a semiclassical approach
\cite{Mahanty72,mah,mah1976}, and was applied to the case of two
atoms placed between two planar, perfectly conducting plates
\cite{mah}. The situation of two atoms between two perfectly
conducting plates was later reconsidered taking into account finite
temperature effects \cite{bos}. Other scenarios such as two atoms
placed within a planar dielectric three-layer geometry \cite{mar} or
two anisotropic molecules in front of a dielectric half space or
within a planar dielectric cavity have also been studied \cite{Cho}.

In this paper we present an exact derivation of a very general formula
for the vdW potential of two ground-state atoms in the presence of an
arbitrary arrangement of dispersing and absorbing magnetodielectric
bodies. Based on macroscopic quantum electrodynamics in linearly,
locally and causally responding media, and starting from the
multipolar coupling Hamiltonian for the atom--field interaction in
electric-dipole approximation, we calculate the vdW potential in
leading, fourth-order perturbation theory. We then apply the
general result to the cases that the two atoms are placed (i) within
bulk material and (ii) in front of a planar magnetodielectric
multilayer system.

The paper is organized as follows. In Sec.~\ref{Ham} the atom--field
interaction Hamiltonian in its multipolar coupling form is presented.
The derivation of the general formula for the vdW potential is given
in Sec.~\ref{p}, and Sec.~\ref{Appls} is devoted to the applications
mentioned, where a detailed analytical as well as numerical
analysis is given. Finally, the paper ends with a summary and
conclusions in Sec.~\ref{con}.


\section{Multipolar-coupling Hamiltonian}
\label{Ham}

The Hamiltonian for a system consisting of nonrelativistic charged
particles $\alpha$ (each particle having charge $q_\alpha$, mass
$m_\alpha$, position $\hat{\mathbf{r}}_\alpha$, and canonically
conjugate momentum $\hat{\mathbf{p}}_\alpha$) interacting with the
electromagnetic field in the presence of dispersing and absorbing
magnetodielectric bodies is given by \cite{Knoll01,Ho03}
\begin{align}
\label{eq1}
\hat{H}=& \sum_{\lambda=e,m}\int\mathrm{d}^3r
 \int_0^{\infty}\mathrm{d}\omega\,
 \hbar\omega\,\hat{\mathbf{f}}_\lambda^\dagger(\mathbf{r},\omega)
 \cdot\hat{\mathbf{f}}_\lambda(\mathbf{r},\omega)\nonumber\\
& +\sum_\alpha\frac{1}{2m_\alpha}
 \Bigl[\hat{\mathbf{p}}_\alpha
 -q_\alpha\hat{\mathbf{A}}(\hat{\mathbf{r}}_\alpha)\Bigr]^2
 +{\textstyle\frac{1}{2}}\int\mathrm{d}^3r\,
 \hat{\rho}_\mathrm{p}(\mathbf{r})\hat{\varphi}_\mathrm{p}
 (\mathbf{r})
 \nonumber\\
& +\int\mathrm{d}^3r\,\hat{\rho}_\mathrm{p}(\mathbf{r})
 \hat{\varphi}(\mathbf{r}),
\end{align}
where
\begin{equation}
\label{eq2}
\hat{\rho}_\mathrm{p}(\mathbf{r})
 =\sum_\alpha q_\alpha\delta(\mathbf{r}-\hat{\mathbf{r}}_\alpha)
\end{equation}
and
\begin{equation}
\label{eq3}
\hat{\varphi}_\mathrm{p}(\mathbf{r})= \int\mathrm{d}^3{r}'\,
 \frac{\hat{\rho}_\mathrm{p}(\mathbf{r}')}
 {4\pi\varepsilon_0|\mathbf{r}-\mathbf{r}'|}
\end{equation}
are the charge density and scalar potential of the particles,
respectively. The Bosonic fields
$\hat{\mathbf{f}}_\lambda(\mathbf{r},\omega)$
and $\hat{\mathbf{f}}_\lambda^\dagger(\mathbf{r},\omega)$ are the
canonically conjugate variables that describe the combined system of
the electromagnetic field and the (inhomogeneous) magnetodielectric
medium, including the dissipative system responsible for absorption,
\begin{align}
\label{eq3.1}
&\Bigl[\hat{f}_{\lambda i}(\mathbf{r},\omega),
 \hat{f}^\dagger_{\lambda' i'}(\mathbf{r}',\omega')\Bigr]
 =\delta_{\lambda\lambda'}\delta_{ii'}\delta(\mathbf{r}-\mathbf{r}')
 \delta(\omega-\omega'),\\[.5ex]
\label{eq3.2}
&\Bigl[\hat{f}_{\lambda i}(\mathbf{r},\omega),
 \hat{f}_{\lambda' i'}(\mathbf{r}',\omega')\Bigr]=0,
\end{align}
where $\lambda$ $\!=$ $\!e$ ($\lambda$ $\!=$ $\!m$) refers to the
electric (magnetic) excitations. The vector potential
$\hat{\mathbf A}({\mathbf r})$ and the scalar potential
$\hat{\varphi}({\mathbf r})$ of the medium-assisted electromagnetic
field can in Coulomb gauge be expressed in terms of the dynamical
variables $\hat{\mathbf{f}}_\lambda(\mathbf{r},\omega)$ and
$\hat{\mathbf{f}}_\lambda^\dagger(\mathbf{r},\omega)$ as
\begin{align}
\label{eq6}
&\hat{\mathbf{A}}(\mathbf{r})=\int_0^\infty\mathrm{d}\omega\,
 (i\omega)^{-1}
 {\underline{\hat{\mathbf{E}}}}{}^\perp(\mathbf{r},\omega)
 +\mathrm{H.c.},
\\[.5ex]
\label{eq7}
&\bm{\nabla}\hat{\varphi}(\mathbf{r})=-\int_0^\infty\mathrm{d}\omega\,
 {\underline{\hat{\mathbf{E}}}}{}^\parallel(\mathbf{r},\omega)
 +\mathrm{H.c.},
\end{align}
with
\begin{equation}
\label{eq8}
\underline{\hat{\mathbf{E}}}(\mathbf{r},\omega)
 =\sum_{\lambda=e,m}\int\mathrm{d}^3r'\,
 \bm{G}_\lambda(\mathbf{r},\mathbf{r}',\omega)
 \cdot\hat{\mathbf{f}}_\lambda(\mathbf{r}',\omega),
\end{equation}
where
\begin{align}
\label{eq9}
&\bm{G}_e(\mathbf{r},\mathbf{r}',\omega)
 =i\,\frac{\omega^2}{c^2}\sqrt{\frac{\hbar}{\pi\varepsilon_0}\,
 \mathrm{Im}\,\varepsilon(\mathbf{r}',\omega)}\,
 \bm{G}(\mathbf{r},\mathbf{r}',\omega),\\
\label{eq10}
&\bm{G}_m({\mathbf r},{\mathbf r}{}',\omega)
 =-i\,\frac{\omega}{c}\,
 \bm{G}(\mathbf{r},\mathbf{r}',\omega)\!\times\!\!
 \overleftarrow{\bm{\nabla}}_{\!\!\mathbf{r}'}
 \sqrt{-\frac{\hbar}{\pi\varepsilon_0}\,
 \mathrm{Im}\,\kappa(\mathbf{r}',\omega)},
\end{align}
$\bigl[\bm{G}(\mathbf{r},\mathbf{r}',\omega)\!\times\!
\overleftarrow{\bm{\nabla}}_{\!\!\mathbf{r}'}\bigr]_{ij}$
$\!=$ $\!\epsilon_{jkl}\partial'_l
G_{ik}(\mathbf{r},\mathbf{r}',\omega)$, and $\perp$ ($\parallel$)
denotes transverse (longitudinal) vector fields. In Eqs.~(\ref{eq9})
and (\ref{eq10}), $\bm{G}(\mathbf{r},\mathbf{r}',\omega)$ is the
classical Green tensor obeying the equation
\begin{equation}
\label{eq14}
\biggl[\bm{\nabla}\times\kappa(\mathbf{r},\omega)\bm{\nabla}\times
 -\frac{\omega^2}{c^2}\,\varepsilon(\mathbf{r},\omega)\biggr]
 \bm{G}(\mathbf{r},\mathbf{r}',\omega)
 =\bm{\delta}(\mathbf{r}-\mathbf{r}')
\end{equation}
together with the boundary condition at infinity. All relevant
characteristics of the macroscopic bodies enter the theory via the
space- and frequency-dependent complex permittivity
$\varepsilon({\mathbf r},\omega)$ and permeability $\mu({\mathbf
r},\omega)$ $\!=$ $\!\kappa^{-1}(\mathbf{r},\omega)$, with the real
and imaginary parts of $\varepsilon({\mathbf r},\omega)$ and
$\kappa(\mathbf{r},\omega)$ satisfying the Kramers--Kronig relations.
Note that the Green tensor obeys the useful properties \cite{Knoll01}
\begin{align}
\label{Gprop1}
&\bm{G}^{\ast}(\mathbf{r},\mathbf{r}',\omega)
 =\bm{G}(\mathbf{r},\mathbf{r}',-\omega^{\ast}),
\\[1ex]
\label{Gprop2}
&\bm{G}(\mathbf{r},\mathbf{r}',\omega)
 =\bm{G}^\top(\mathbf{r}',\mathbf{r},\omega),
\\[1ex]
\label{Gprop3}
& \sum_{\lambda=e,m}\int\mathrm{d}^3 s\,
 \bm{G}_\lambda(\mathbf{r},\mathbf{s},\omega)\cdot
\bm{G}^+_\lambda(\mathbf{r}',\mathbf{s},\omega)
\nonumber\\
&\hspace{5ex}
 =\frac{\hbar\mu_0}{\pi}\,\omega^2
 \mathrm{Im}\,\bm{G}(\mathbf{r},\mathbf{r}',\omega).
\end{align}

If the charged particles constitute a system of neutral atoms and/or
molecules (briefly referred to as atoms in the following) labelled by
$A$, $\sum_{\alpha\in A}q_\alpha$ $\!=$ $\!0$, then it is convenient
to employ the Hamiltonian in the multipolar-coupling form, which can
be obtained from the minimal-coupling form (\ref{eq1}) via a
Power--Zienau transformation
\begin{equation}
\label{PZ}
\hat{U}=\exp\biggl[\frac{i}{\hbar}\int\mathrm{d}^3r\,
  \sum_A \hat{\mathbf{P}}_A(\mathbf{r})\cdot
  \hat{\mathbf{A}}(\mathbf{r})\biggr],
\end{equation}
where the polarization of atom $A$ is given by
\begin{equation}
\label{polA}
\hat{\mathbf{P}}_A(\mathbf{r})
 =\sum_{\alpha\in A}q_\alpha\hat{\bar{\mathbf{r}}}_\alpha
 \int _0^1\mathrm{d}\lambda\,
 \delta(\mathbf{r}-\hat{\mathbf{r}}_A
 -\lambda\hat{\bar{\mathbf{r}}}_\alpha),
\end{equation}
with
\begin{equation}
\label{eq27}
\hat{\bar{\mathbf{r}}}_\alpha
 =\hat{\mathbf{r}}_\alpha-\hat{\mathbf{r}}_A
\end{equation}
denoting the particle coordinates relative to the center of mass
\begin{equation}
\label{eq14.1}
\hat{\mathbf{r}}_A=\sum_{\alpha\in A}
 \frac{m_\alpha}{m_A}\,\hat{\mathbf{r}}_\alpha
\end{equation}
of atom $A$ ($m_A$ $\!=$ $\!\sum_{\alpha\in A} m_\alpha$). We assume
that all the atoms are (i) essentially at rest, $m_\alpha/m_A$ $\!\to$
$\!0$, (ii) small compared to the wavelength of the relevant field
components, $\hat{\bar{\mathbf{r}}}_\alpha$ $\!\to$
$\!\hat{\mathbf{r}}_A$, and (iii) well separated from each other,
\begin{equation}
\label{eq14.2}
\int\mathrm{d}^3r\,\hat{\mathbf{P}}_A(\mathbf{r})
 \cdot\hat{\mathbf{P}}_{B}(\mathbf{r})
 =\delta_{AB}\int\mathrm{d}^3r\,
 \hat{\mathbf{P}}^2_{A}(\mathbf{r}).
\end{equation}
Under these assumptions, the Hamiltonian in the multipolar coupling
scheme can be obtained from Eqs.~(\ref{eq1}) and (\ref{PZ}) in
complete analogy to the procedure outlined in Ref.~\cite{Buhmann04},
resulting in
\begin{equation}
\label{eq43}
\hat{H}=\hat{H}_\mathrm{F}+\sum_A\hat{H}_A+\sum_A
\hat{H}_{A\mathrm{F}},
\end{equation}
where
\begin{align}
\label{eq44}
&
\hat{H}_\mathrm{F}
= \sum_{\lambda=e,m}\int\mathrm{d}^3r\,
 \int_0^{\infty}\mathrm{d}\omega\,\hbar\omega\,
 \hat{\mathbf{f}}_\lambda{\!^\dagger}(\mathbf{r},\omega)
 \cdot\hat{\mathbf{f}}_\lambda(\mathbf{r},\omega),
\\[1ex]
\label{eq45}
&
\hat{H}_A
=\sum_{\alpha\in A}
 \frac{\hat{\mathbf{p}}_\alpha{\!^2}}{2m_\alpha}
 +\frac{1}{2\varepsilon_0}\int\mathrm{d}^3r\,
 \hat{\mathbf{P}}_A^2(\mathbf{r}),
\\[1ex]
\label{eq46}
&
\hat{H}_{A\mathrm{F}}
= -\hat{\mathbf{d}}_A\cdot
 \hat{\mathbf{E}}(\hat{\mathbf{r}}_{A})
 \!+\!\sum_{\alpha\in A}\frac{q_\alpha}{2m_\alpha}
\hat{\bar{\mathbf{p}}}_\alpha\cdot[\hat{\bar{\mathbf{r}}}_\alpha
 \times\hat{\mathbf{B}}(\hat{\mathbf{r}}_{A})]
 \nonumber\\
&\hspace{7ex}
+\sum_{\alpha\in A}\frac{q_\alpha^2}{8m_\alpha}
 \bigl[\hat{\bar{\mathbf{r}}}_\alpha
 \times\hat{\mathbf{B}}(\hat{\mathbf{r}}_{A})\bigr]^2.
\end{align}
In Eq.~(\ref{eq46}),
\begin{equation}
\label{eq34}
\hat{\mathbf{d}}_A=\sum_{\alpha\in A}
 q_\alpha\hat{\bar{\mathbf{r}}}_\alpha
 =\sum_{\alpha\in A}
 q_\alpha\hat{\mathbf{r}}_\alpha
\end{equation}
is the electric dipole moment of atom $A$, and the electric and
induction fields are given by
\begin{equation}
\label{eq20}
\hat{\mathbf{E}}(\mathbf{r})=\int_0^\infty\mathrm{d}\omega\,
 \underline{\hat{\mathbf{E}}}(\mathbf{r},\omega)+\mathrm{H.c.},
\end{equation}
with $\underline{\hat{\mathbf{E}}}(\mathbf{r},\omega)$
from Eq.~(\ref{eq8}), and
\begin{align}
\label{eq21}
&\hat{\mathbf{B}}(\mathbf{r})=\int_0^\infty\mathrm{d}\omega\,
 \underline{\hat{\mathbf{B}}}(\mathbf{r},\omega)+\mathrm{H.c.},\\
\label{eq22}
&\underline{\hat{\mathbf{B}}}(\mathbf{r},\omega)
 =(i\omega)^{-1}\bm{\nabla}\times
 \underline{\hat{\mathbf{E}}}(\mathbf{r},\omega).
\end{align}
Note that in the multipolar-coupling scheme
$\hat{\mathbf{E}}(\mathbf{r})$ has the physical meaning of a
displacement field w.r.t. the polarization of the atoms.
Finally, in the case of atoms which are not magnetically polarizable,
we may omit the second and third terms in Eq.~(\ref{eq46}) so that
Eq.~(\ref{eq46}) reduces to the well-known electric-dipole term
\begin{equation}
\label{eq49}
\hat{H}_{A\mathrm{F}}
=-\hat{\mathbf{d}}_A\cdot\hat{\mathbf{E}}(\hat{\mathbf{r}}_A).
\end{equation}


\section{The van der Waals potential}
\label{p}

Let us consider two neutral, ground-state atoms $A$ and $B$
at given positions $\mathbf{r}_A$ and $\mathbf{r}_B$
in the presence of arbitrarily shaped magnetodielectric bodies.
Denoting by $|n_{A(B)}\rangle$ the (unperturbed) energy eigenstates of
atom $A(B)$, we may represent the atomic Hamiltonian $H_{A(B)}$,
Eq.~(\ref{eq45}), in the form
 \begin{equation}
 \label{Ha}
 \hat{H}_{A(B)}=
 \sum_{n}E^n_{A(B)}
 |n_{A(B)}\rangle\langle n_{A(B)}|.
 \end{equation}
Restricting our attention to the electric-dipole approximation,
the interaction Hamiltonian $\hat{H}_{A(B)\mathrm{F}}$ reads,
according to Eq.~(\ref{eq49}) [$\hat{\mathbf{r}}_{A(B)}
\mapsto\mathbf{r}_{A(B)}$],
\begin{multline}
\label{eq49-1}
 \hat{H}_{A(B)\mathrm{F}} = \\
 -\sum_n\sum_m
 |n_{A(B)}\rangle\langle m_{A(B)}|
 \mathbf{d}_{A(B)}^{nm}
 \cdot \hat{\mathbf{E}}(\mathbf{r}_{A(B)}),
\end{multline}
where
${\mathbf d}_{A(B)}^{nm}$ $\!=$ $\!\langle n_{A(B)}|
\hat{\mathbf{d}}_{A(B)}|m_{A(B)}\rangle$, and
$\hat{\mathbf{E}}(\mathbf{r})$ is given by Eq.~(\ref{eq20}) together
with Eq.~(\ref{eq8}). Further, let $|\{0\}\rangle$,
$|1^{(\alpha)}\rangle$, and $|1^{(\beta)},1^{(\gamma)}\rangle$ be the
vacuum, single-, and two-quantum excited states of the combined system
consisting of the electromagnetic field and the bodies, respectively,
\begin{align}
\label{gl31}
&{\hat f}_{\lambda i}(\mathbf r,\omega)|\{0\}\rangle=0,\\[.5ex]
\label{gl32}
&\hat{f}_{\lambda_\alpha
 i_\alpha}^\dagger(\mathbf{r_\alpha},\omega_\alpha)|\{0\}\rangle
 \equiv |1^{(\alpha)}\rangle,\\[.5ex]
\label{gl33}
&{\textstyle\frac{1}{\sqrt{2}}}\,
 \hat{f}_{\lambda_\beta i_\beta}^\dagger
 (\mathbf{r_\beta},\omega_\beta)
 \hat{f}_{\lambda_\gamma i_\gamma}^\dagger
 (\mathbf{r_{\gamma}},\omega_\gamma)|\{0\}\rangle
 \equiv |1^{(\beta)},1^{(\gamma)}\rangle
\end{align}
[the corresponding single- und two-excitation energies are
respectively $\hbar\omega_\alpha$ and
$\hbar(\omega_\beta+\omega_\gamma)$].

Following Casimir's and Polder's approach \cite{c-p}
(see also Ref.~\cite{Craig84}), we identify the two-atom vdW
interaction with the position-dependent shift of the ground-state
energy $\Delta E_{AB}$ calculated in leading-order perturbation theory
according to
\begin{align}
\label{E12}
&\hspace{-2ex}
\Delta E_{AB}=-
\hspace{-1ex}
\sideset{}{'}\sum_{I,II,III}
\hspace{-1ex}
 \frac{\langle 0|\hat{H}_{A\mathrm{F}}
 \!+\!\hat{H}_{B\mathrm{F}}|III\rangle
 \langle III|\hat{H}_{A\mathrm{F}}
 \!+\!\hat{H}_{B\mathrm{F}}|II\rangle}
 {(E_{I}-E_0)}\nonumber\\
&\hspace{10ex}\times\frac{\langle II|\hat{H}_{A\mathrm{F}}
 \!+\!\hat{H}_{B\mathrm{F}}|I\rangle
 \langle I|\hat{H}_{A\mathrm{F}}
 \!+\!\hat{H}_{B\mathrm{F}}|0\rangle}
 {(E_{II}-E_0)(E_{III}-E_0)}\,,
\end{align}
where the primed sum indicates that only intermediate
states $|I\rangle$, $|II\rangle$, and $|III\rangle$
other than the (unperturbed) ground state of the overall system,
\begin{equation}
\label{eq25}
 |0\rangle = |0_A\rangle |0_B\rangle
 |\{0\}\rangle,
\end{equation}
are included in the summations. Note that the summations
include position and frequency integrals.

From Eq.~(\ref{eq49-1}), by considering only two-atom virtual
processes, it can be inferred that the intermediate states $|I\rangle$
and $|III\rangle$ have one of the atoms excited and one body-assisted
field excitation present, while the intermediate states $|II\rangle$
can be of three types: (i) both atoms in the ground state with two
field excitations present, (ii) both atoms excited with no field
excitation present, and (iii) both atoms excited with two field
excitations present. All possible intermediate states together with
the respective energy denominators are listed in Tab.~\ref{denom1} in
App.~\ref{mat-elm}. 

Let us consider, e.g., case (1) in
this table. Substituting the corresponding matrix elements
(\ref{mel1})--(\ref{mel4}) as given in App.~\ref{mat-elm} into
Eq.~(\ref{E12}), we derive the contribution $\Delta E_{AB(1)}$ to the
two-atom energy shift $\Delta E_{AB}$ to be
\begin{widetext}
\begin{align}
\label{wow}
 &\Delta E_{AB(1)}=-\frac{1}{2\hbar^3}
 \sum_{n,m}\,\sum_{i_1,i_2,i_3,i_4}
 \sum_{\lambda_1,\lambda_2,\lambda_3,\lambda_4}
 \Biggl[\prod_{j=1}^4\int\mathrm{d}^3r_j
 \int_0^\infty\mathrm{d}\omega_j\Biggr]
 \,\frac{1}{D_{nm}(\omega_1,\omega_2,\omega_3,\omega_4)}\nonumber\\
&\times\Big\{ \big[\mathbf d_A^{n0}\!\cdot\!
 \bm G_{\lambda_1}^\ast(\mathbf r_{A},\mathbf r_1,\omega_1)\big]_{i_1}
 \big[\mathbf d_A^{0n}\!\cdot\!\bm G_{\lambda_3}^\ast
 (\mathbf r_{A},\mathbf r_3,\omega_3)\big]_{i_3}
 \big[\mathbf d_B^{m0}\!\cdot\!\bm G_{\lambda_3}
 (\mathbf r_{B},\mathbf r_3,\omega_3)\big]_{i_3}
 \big[\mathbf d_B^{0m}\!\cdot\!\bm G_{\lambda_4}
 (\mathbf r_{B},\mathbf r_4,\omega_4)\big]_{i_4}
 \delta^{(12)}\delta^{(24)}\nonumber\\
&\quad + \big[\mathbf d_A^{n0}\!\cdot\!
 \bm G_{\lambda_1}^\ast(\mathbf r_{A},\mathbf r_1,\omega_1)\big]_{i_1}
 \big[\mathbf d_A^{0n}\!\cdot\!\bm G_{\lambda_3}^\ast
 (\mathbf r_{A},\mathbf r_3,\omega_3)\big]_{i_3}
 \big[\mathbf d_B^{m0}\!\cdot\!\bm G_{\lambda_2}
 (\mathbf r_{B},\mathbf r_2,\omega_2)\big]_{i_2}
 \big[\mathbf d_B^{0m}\!\cdot\!\bm G_{\lambda_4}
 (\mathbf r_{B},\mathbf r_4,\omega_4)\big]_{i_4}
 \delta^{(12)}\delta^{(34)}\nonumber \\
&\quad + \big[\mathbf d_A^{n0}\!\cdot\!
 \bm G_{\lambda_1}^\ast(\mathbf r_{A},\mathbf r_1,\omega_1)\big]_{i_1}
 \big[\mathbf d_A^{0n}\!\cdot\!\bm G_{\lambda_2}^\ast
 (\mathbf r_{A},\mathbf r_2,\omega_2)\big]_{i_2}
 \big[\mathbf d_B^{m0}\!\cdot\!\bm G_{\lambda_2}
 (\mathbf r_{B},\mathbf r_2,\omega_2)\big]_{i_2}
 \big[\mathbf d_B^{0m}\!\cdot\!\bm G_{\lambda_4}
 (\mathbf r_{B},\mathbf r_4,\omega_4)\big]_{i_4}
 \delta^{(13)}\delta^{(34)}\nonumber \\
&\quad + \big[\mathbf d_A^{n0}\!\cdot\!
 \bm G_{\lambda_1}^\ast(\mathbf r_{A},\mathbf r_1,\omega_1)\big]_{i_1}
 \big[\mathbf d_A^{0n}\!\cdot\!\bm G_{\lambda_2}^\ast
 (\mathbf r_{A},\mathbf r_2,\omega_2)\big]_{i_2}
 \big[\mathbf d_B^{m0}\!\cdot\!\bm G_{\lambda_3}(
 \mathbf r_{B},\mathbf r_3,\omega_3)\big]_{i_3}
 \big[\mathbf d_B^{0m}\!\cdot\!\bm G_{\lambda_4}
 (\mathbf r_{B},\mathbf r_4,\omega_4)\big]_{i_4}
 \delta^{(13)}\delta^{(24)}\Big\},
\end{align}
\end{widetext}
where
\begin{equation}
\label{delta}
\delta^{(\alpha\beta)}
 =\delta_{i_\alpha i_\beta}\delta_{\lambda_\alpha \lambda_\beta}
 \delta(\mathbf r_\alpha-\mathbf r_\beta)
 \delta(\omega_\alpha-\omega_\beta)
\end{equation}
and
\begin{align}
\label{den}
D_{nm}(\omega_1,\omega_2,\omega_3,\omega_4) =
(\omega_A^n\!+\!\omega_1)(\omega_2\!+\!\omega_3)
(\omega_B^m\!+\!\omega_4)
\end{align}
[$\omega_{A(B)}^n$ $\!=$ $\!(E^n_{A(B)}$ $\!-$ $\!E^0_{A(B)})/\hbar$].
Recalling Eq.~(\ref{Gprop3}), we may simplify Eq.~(\ref{wow}) to
\begin{align}
\label{E20}
& \Delta E_{AB(1)}= - \frac{\mu_0^2}{\hbar\pi^2}
     \sum_{n,m}
     \int_0^\infty {\mathrm d}\omega \int_0^\infty {\mathrm d}
     \omega'\,\omega^2\omega'^{2}
\nonumber\\
&\quad
\times
     \bigg(\frac{1}{D_{\mathrm i}}+\frac{1}{D_{\mathrm {ii}}}\bigg)
     \big[{\mathbf d}_A^{0n}\!\cdot\!
     {\rm Im}\bm{G}({{\mathbf r}_A},
     {{\mathbf r}_{B}},\omega)
     \!\cdot\!{\mathbf d}_B^{0m}\big]
\nonumber\\
&\quad\hspace{10ex}
\times
     \big[{\mathbf d}_A^{0n}\!\cdot\!
     {\rm Im}\bm{G}({{\mathbf r}_{A}},
     {{\mathbf r}_{B}},\omega')
     \!\cdot\!{\mathbf d}_{B}^{0m}\big],
\end{align}
where $D_{\mathrm i}$ and $D_{\mathrm {ii}}$ are respectively
the first and the second denominators in Tab.~\ref{denom1},
and without loss of generality we have assumed that the matrix
elements of the electric-dipole operators are real.

The contributions $\Delta E_{AB(k)}$ to $\Delta E_{AB}$ which
correspond to the cases (2)--(10) in Table~\ref{denom1}
in App.~\ref{mat-elm} can be calculated analogously. It turns out that
they differ from Eq.~(\ref{E20}) only in the energy denominators. It
is not difficult to prove that summation of the energy denominators
under the double frequency integral leads to (App.~\ref{AppA})
\begin{multline}
\label{E21}
     \sum_{ a=\mathrm i}^{\mathrm {xii}} \frac{1}{D_a}\to
        \frac{4(\omega_A^n+\omega_B^m+\omega)}
     {(\omega_A^n+\omega_B^m)
     (\omega_A^n+
     \omega)
 (\omega_{B}^{m}+\omega)}\\
 \times\left( \frac{1}{\omega+\omega'} - \frac{1}{\omega-\omega'}
 \right).
\end{multline}
Hence, the two-atom contributions $\Delta E_{AB(k)}$ to the
fourth-order energy shift lead to the vdW potential
$U_{AB}(\mathbf{r}_A,\mathbf{r}_B)$ $\!=$ $\!\sum_{k=1}^{10}
\Delta E_{AB(k)}$ as follows:
\begin{align}
\label{E22}
&U_{AB}(\mathbf{r}_A,\mathbf{r}_B)
      = - \frac{4\mu_0^2}{\hbar\pi^2}
         \sum_{n,m} \frac{1}{\omega_A^n+
         \omega_B^m}
     \int_0^\infty {\rm d}\omega \int_0^\infty {\rm d}\omega'
\nonumber\\
&\times\frac{\omega^2\omega'^{2}(\omega_A^n+
     \omega_B^m+
     \omega)}
     {(\omega_A^n+\omega)
     (\omega_B^m+\omega)}
       \left( \frac{1}{\omega+\omega'} - \frac{1}{\omega-\omega'}
\right)\nonumber\\[1ex]
&\times[{\mathbf d}_A^{0n}\!\cdot\!
     {\rm Im}\bm{G}({{\mathbf r}_{A}},
     {{\mathbf r}_{B}},\omega)
      \!\cdot\!{\mathbf d}_B^{0m}]
     [{\mathbf d}_A^{0n}\!\cdot\!
     {\rm Im}\bm{G}({{\mathbf r}_{A}},
     {{\mathbf r}_{B}},\omega')\!\cdot\!
     {\mathbf d}_B^{0m}].
\end{align}

To perform the integral over $\omega'$, we first
use the identity ${\rm Im}\,\bm{G}$ $\!=$ $\!(\bm {G}$ $\!-$
$\!\bm{G}^\ast)/(2i)$ and the relation (\ref{Gprop1}) to write
\begin{multline}
\label{E23}
\hspace{-2ex}
     \int_0^\infty {\rm d}\omega'
     \left( \frac{1}{\omega+\omega'} - \frac{1}{\omega-\omega'}
\right)
     \omega'^2 {\rm Im}\bm{G}({{\mathbf r}_{A}},
     {{\mathbf r}_{B}},\omega')\\
     =\frac{1}{2i} \int_{-\infty}^{\infty} {\rm d}\omega'
     \left( \frac{1}{\omega+\omega'} - \frac{1}{\omega-\omega'}
\right)
     \omega'^2 \bm{G}({{\mathbf r}_{A}},
     {{\mathbf r}_{B}},\omega'),
\end{multline}
where the poles at $\omega'$ $\!=$ $\!-\omega$ and $\omega'$ $\!=$
$\!\omega$ are to be treated as principal values. The Green tensor
is analytic in the upper half of the complex frequency plane including
the real axis, apart from a possible pole at the origin. In addition,
$\omega'^2 \bm{G} ({{\mathbf r}_{A}},{{\mathbf r}_{B}},\omega')$ is
well-behaved for vanishing $\omega'$ \cite{Knoll01}. We may therefore
replace the integral on the right hand side of Eq.~(\ref{E23}) by
contour integrals along infinitely small half-circles surrounding
$\pm\omega$, and an infinitely large half-circle in the upper complex
half-plane. The integral along the infinitely large half-circle
vanishes because \cite{Knoll01}
\begin{equation}
\label{Glim}
     \lim_{|\omega|\rightarrow \infty} \omega^2
     \bm{G}({\mathbf r}_{A},{\mathbf r}_{B},\omega)
     \Big|_{{\mathbf r}_{A}\neq{\mathbf r}_{B}} = 0.
\end{equation}
Collecting the contributions from the infinitely small half-circles,
we end up with
\begin{multline}
\label{E25}
     \int_0^\infty {\rm d}\omega'
     \left( \frac{1}{\omega+\omega'} - \frac{1}{\omega-\omega'}
\right)
     \omega'^2 {\rm Im}\bm{G}({\mathbf r}_{A},
     {\mathbf r}_{B},\omega')\\
     ={\textstyle\frac{1}{2}} \pi \omega^2
     [\bm{G}({\mathbf r}_{A},{\mathbf r}_{B},\omega)+
      \bm{G}^\ast({\mathbf r}_{A},
      {\mathbf r}_{B},\omega)],
\end{multline}
where we have again made use of the relation (\ref{Gprop1}).
Substitution of Eq.~(\ref{E25}) into Eq.~(\ref{E22}) leads to
\begin{widetext}
\begin{align}
\label{E26}
&U_{AB}(\mathbf{r}_A,\mathbf{r}_B)
       = -
     \frac{\mu_0^2}{i\hbar\pi}
          \sum_{n,m}
     \frac{1}{\omega_A^n\!+\!\omega_B^m}
     \int_0^\infty\!\!\!\! {\rm d}\omega\,
     \frac{\omega^4 (\omega_A^n\!+\!\omega_B^m\!+\!
     \omega)}
     {(\omega_A^n\!+\!\omega)
     (\omega_B^m\!+\!\omega)}
 \Bigl\{
     [{\mathbf d}_A^{0n}\!\cdot\!
     \bm{G}({{\mathbf r}_{A}},{{\mathbf r}_{B}},
     \omega)
     \!\cdot\!  {\mathbf d}_B^{0m}]^2
     - [{\mathbf d}_A^{0n}\!\cdot\!
     \bm{G}^\ast({{\mathbf r}_{A}},{{\mathbf r}_{B}},
     \omega) \!\cdot\! {\mathbf d}_B^{0m}]^2
          \Bigr\}
\nonumber\\
&\quad
     = -
     \frac{\mu_0^2}{i\hbar\pi}
          \sideset{}{'}\sum_{n,m}
    \frac{1}{\omega_A^n\!+\!\omega_B^m} \biggl\{
     \int_0^\infty {\rm d}\omega
     \frac{\omega^4 (\omega_A^n\!+\!\omega_B^m
     \!+\!\omega)}
     {(\omega_A^n\!+\!\omega) (\omega_B^m\!+\!
     \omega)}
+ \int_0^{-\infty} {\rm d}\omega
     \frac{\omega^4 (\omega_A^n+
     \omega_B^m-\omega)}
     {(\omega_A^n-\omega) (\omega_B^m-
     \omega)}\biggr\}
     [{\mathbf d}_A^{0n}\!\cdot\!
     \bm{G}({{\mathbf r}_{A}},{{\mathbf r}_{B}},
     \omega)
      \!\cdot\! {\mathbf d}_B^{0m}]^2.
\end{align}
\end{widetext}

This equation can be further simplified by again using
contour-integral techniques. It can be seen that the integrand
in the first integral in Eq.~(\ref{E26}) is analytic in the first
quadrant of the complex frequency plane, including the positive real
axis. Therefore, it can be replaced by contour integrals along an
infinitely large quarter-circle in the first quadrant and along the
positive imaginary axis, introducing a purely imaginary frequency,
$\omega$ $\!=$ $\!iu$. The integral along the infinitely large
quarter-circle vanishes because of Eq.~(\ref{Glim}). In a similar way,
the second integral in Eq.~(\ref{E26}) can also be transformed to one
over the imaginary axis. Combining the contributions from the two
integrals leads to
\begin{multline}
\label{E27}
U_{AB}(\mathbf{r}_A,\mathbf{r}_B)
 = - \frac{2\mu_0^2}{\hbar\pi}
     \sum_{n,m}
     \int_0^\infty\!\!\!\!
     \frac{{\rm d}u\,u^4\omega_A^n\omega_B^m
     }{[(\omega_A^n)^2\!+\!u^2][(\omega_B^m)^2\!+\!u^2]}\\
\times\,
     [{\mathbf d}_A^{0n}\!\cdot\!
     \bm{G}({{\mathbf r}_{A}},{{\mathbf r}_{B}},iu)
     \!\cdot\! {\mathbf
     d}_B^{0m}]^2.
\end{multline}
An expression of this type was first given in Ref.~\cite{mah1976} on
the basis of a heuristic generalization of the respective free-space
result.

Noting that the (lowest-order) atomic ground-state polarizability
tensor is (see, e.g., \cite{Fain63})
\begin{align}
\label{atpol1}
     \bm{\alpha}_{A(B)}(\omega)=& \lim_{\eta\to 0+}
     \frac{2}{\hbar}\sum_n
     \frac{ \omega_{A(B)}^{n}{\mathbf d}_{A(B)}^{0n}
      {\mathbf d}_{A(B)}^{n0}}
      {(\omega_{A(B)}^n)^2-\omega^2-i\eta\omega}\,,
\end{align}
we may rewrite Eq.~(\ref{E27}) as
\begin{multline}
\label{E29}
U_{AB}(\mathbf{r}_A,\mathbf{r}_B)
=-     \frac{\hbar\mu_0^2}{2\pi}
     \int_0^{\infty} {\rm d}u \,u^4\\
\times
     \mathrm{Tr}\bigl[\bm{\alpha}_A(iu)\!\cdot\!
     \bm{G}(\mathbf{r}_A,\mathbf{r}_B,iu)
\cdot\!\bm{\alpha}_B(iu)\!\cdot\!
   \bm{G}(\mathbf{r}_B,\mathbf{r}_A,iu)\bigr],
\end{multline}
where we have used Eq.~(\ref{Gprop2}).
In particular for atoms, which are spherically symmetric,
\begin{align}
\label{atpol2}
\bm{\alpha}_{A(B)}(\omega)
&=\alpha_{A(B)}(\omega)\bm{I}\nonumber\\
&=\lim_{\eta\to 0+} \frac{2}{3\hbar}\sum_n
 \frac{\omega_{A(B)}^n|{\mathbf d}_{A(B)}^{0n}|^2\bm{I}}
 {(\omega_{A(B)}^n)^2-\omega^2-i\eta\omega}\,,
\end{align}
Eq.~(\ref{E29}) becomes
\begin{multline}
\label{E31}
U_{AB}(\mathbf{r}_A,\mathbf{r}_B)
 =-\frac{\hbar\mu_0^2}{2\pi}\int_0^\infty\mathrm{d}u\,u^4
 \alpha_{A}(iu)\alpha_{B}(iu)\\
 \times\,\mathrm{Tr}\bigl[
 \bm{G}(\mathbf{r}_A,\mathbf{r}_B,iu)
 \!\cdot\!\bm{G}(\mathbf{r}_B,\mathbf{r}_A,iu)
 \bigr].
\end{multline}

The total force acting on atom $A$ and $B$ can be derived from the
potential
\begin{equation}
\label{gl51}
U(\mathbf{r}_A,\mathbf{r}_B)
 =U_{A}(\mathbf r_{A}) +U_B(\mathbf r_B)+
 U_{AB}(\mathbf r_A,\mathbf r_B)
\end{equation}
according to
\begin{equation}
\label{eq66}
\mathbf{F}_{A(B)}
  =-\bm{\nabla}_{\mathbf{r}_{A(B)}}
  U(\mathbf{r}_A,\mathbf{r}_B),
\end{equation}
where $U_{A(B)}$ is the single-atom potential (see, e.g.,
Ref.~\cite{Buhmann04})
\begin{multline}
\label{gl53}
U_{A(B)}(\mathbf{r}_{A(B)})
= \frac{\hbar\mu_0}{2\pi}
\int_0^\infty \mathrm{d}u\, u^2
\\
\times\,\mathrm{Tr}\bigl[\bm{\alpha}_{A(B)}(iu)
\cdot\bm{G}^{(1)}(\mathbf{r}_{A(B)},\mathbf r_{A(B)},iu)\bigr],
\end{multline}
with $\bm{G}^{(1)}$ being the scattering part
of the Green tensor,
\begin{equation}
\label{gl64}
\bm{G}(\mathbf{r},\mathbf{r}',iu)
= \bm{G}^{(0)}(\mathbf{r},\mathbf{r}',iu)
+ \bm{G}^{(1)}(\mathbf{r},\mathbf{r}',iu)
\end{equation}
[$\bm{G}^{(0)}$, bulk part]. In particular, the body-assisted force
acting on atom $A(B)$ due to the presence of atom $B(A)$ reads
\begin{equation}
\label{inforce}
\mathbf F_{AB(BA)}=-\bm \nabla_{\!\mathbf r_{A(B)}}U_{AB}(\mathbf
r_A,\mathbf r_B).
\end{equation}
Note that $\mathbf{F}_{AB}$ $\!\neq$ $-\mathbf{F}_{BA}$ in general,
due to the presence of the bodies.


\section{Applications}
\label{Appls}

\subsection{Bulk material}
\label{Bulk}

Let us first consider the simplest configuration where the two atoms
are embedded in a bulk magnetodielectric
material whose Green tensor reads \cite{Ho03}
\begin{align}
&\bm G(\mathbf r ,\mathbf r',iu)=
\bm G^{(0)}(\mathbf r ,\mathbf r',iu)
\nonumber\\&
\quad=
\frac{\mu(iu)}{4\pi |\mathbf{r}-\mathbf{r}'|}
\left[f(\xi)\bm{I}-g(\xi)
\frac{(\mathbf{r}\!-\!\mathbf{r}')(\mathbf{r}\!-\!\mathbf{r}')}
{|\mathbf{r}-\mathbf{r}'|^2}\right]
\nonumber\\&\hspace{30ex}\times
e^{-n(iu)|\mathbf{r}-\mathbf{r}'|u/c},
\label{bg}
\end{align}
where $\!n(iu)$ $\!=$ $\!\sqrt{\varepsilon(iu)\mu(iu)}\,$ 
and
\begin{align}
\label{f}
&f(x)=1+x+x^2,\\
\label{g}
&g(x)=1+3x+3x^2,\\
\label{xi}
&\xi=c [n(iu)|\mathbf{r}-\mathbf{r}'|u]^{-1}.
\end{align}
Combining Eq.~(\ref{gl51}) [together with Eqs.~(\ref{E31}) and
(\ref{gl53})] with Eq.~(\ref{bg}), we find that ($l$ $\!=$
$\!|\mathbf{r}_A$ $\!-$ $\!\mathbf{r}_B|$)
\begin{align}
\label{Bpot}
&      U(\mathbf r_{A},\mathbf r_{B}) =
 U_{AB}(\mathbf r_{A},\mathbf r_{B})
 \nonumber\\
&\hspace{1ex}=-\frac{\hbar\mu_0^2}{16\pi^3 l^6}
 \int_0^\infty {\rm d} u\,
 \frac{\alpha_{A}(iu)\alpha_{B}(iu)}{\varepsilon^2(iu)}
 e^{-2n(iu)ul/c}\nonumber\\
&\hspace{12ex}\times
 \Bigl\{3 + 6n(iu)ul/c + 5[n(iu)ul/c]^2
 \nonumber\\&\hspace{17ex}
 + 2[n(iu)ul/c]^3+[n(iu)ul/c]^4\Bigr\},
\end{align}
which generalizes earlier results \cite{c-p} on the two-atom vdW
interaction in free space. Note that in Eq.~(\ref{Bpot}) local-field
corrections are disregarded. They could be taken into account in a
similar way as in the case of single-atom systems (see, e.g.,
Ref.~\cite{Scheel99,Ho03,Ho06}).

In the retarded limit, where $l$ $\!\gg$ $\! c/\omega_{\mathrm{min}}$
[$\omega_{\mathrm{min}}$ $\!=$ $\!{\mathrm{min}}(\{\omega_{A'}^n,
\omega_\nu |A'\!=\!A,B;\,n,\nu\!=\!1,2,\ldots\})$, with $\omega_\nu$
denoting the resonance frequencies of the medium], due to the presence
of the exponential in the integrand in Eq.~(\ref{Bpot}), only small
values of $u$ significantly contribute. Hence we may approximately
replace the atomic polarizabilities and the permittivity and
permeability of the medium by their respective static values,
\begin{equation}
\label{alpha0}
\alpha_{A(B)}(iu)\simeq\alpha_{A(B)}(0),\quad\varepsilon(iu)
\simeq\varepsilon(0),\quad\mu(iu)\simeq\mu(0),
\end{equation}
and perform the integral in closed form to yield
\begin{equation}
\label{E35}
      U(\mathbf r_{A},\mathbf r_{B}) = -\frac{C_\mathrm{r}}{l^7}\,,
\end{equation}
where
\begin{equation}
\label{cr}
C_\mathrm{r}=\frac{23\hbar c }{64\pi^3\varepsilon_0^2}\,
\frac{\alpha_{A}(0)\alpha_{B}(0)}{n(0)\varepsilon^2(0)}\,.
 \end{equation}
Equation (\ref{E35}) reveals that the potential behaves like $l^{-7}$
just as in the free-space case, but with the coefficient being reduced
by a factor of $[n(0)\varepsilon^2(0)]^{-1}$.

In the nonretarded limit, where \mbox{$l$ $\!\ll$
$\!c/[n(0)\omega_\mathrm{max}]$} [$\omega_{\mathrm{max}}$ $\!=$
$\!{\mathrm{max}}(\{\omega_{A'}^n,\omega_\nu|A'\!=\!A,B;\,
n,\nu\!=\!1,2,\ldots\})$], the integral in Eq.~(\ref{Bpot}) is
effectively limited to a region where $e^{-2n(iu)ul/c}$ $\!\simeq$
$\!1$ and the term in curly brackets is approximately equal to 3, so
that
\begin{equation}
\label{E36}
      U(\mathbf r_{A},\mathbf r_{B}) =-\frac{C_{\mathrm{nr}}}{l^6}\,,
\end{equation}
where
\begin{equation}
\label{cnr}
C_{\mathrm{nr}}=\frac{3\hbar}{16\pi^3\varepsilon_0^2}
      \int_0^\infty {\rm d}u\,
      \frac{\alpha_{A}(iu)
      \alpha_{B}(iu)}{\varepsilon^2(iu)}\,,
\end{equation}
which shows the $l^{-6}$-dependence also known from the free-space
case. According to Eq.~(\ref{Bpot}) and Eqs.~(\ref{E35})--(\ref{cnr}),
a bulk magnetodielectric medium tends to inhibit the interaction
between the atoms, thereby reducing the interatomic dispersion force.


\subsection{Multilayer systems}
\label{multy}

\begin{figure}
\noindent
\begin{center}
\includegraphics[width=.9\linewidth]{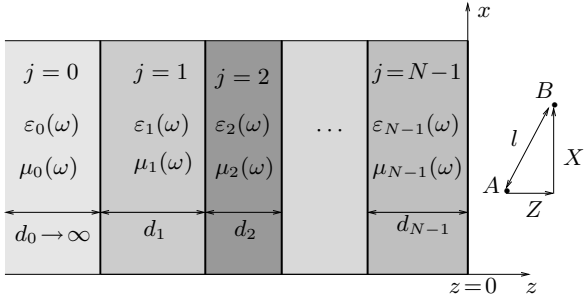}
\end{center}
\caption{
\label{multilayer}
Sketch of the planar multilayer medium.}
\end{figure}

Now let the two atoms be in front of a planar magnetodielectric
multilayer system consisting of $N$ adjoined layers labeled by $j$
($j$ $\!=$ $\!0,1,2,..,N$ $\!-$ $\!1)$ with thicknesses $d_j$ ($d_0$
$\!\to$ $\!\infty$),  permittivities $\varepsilon_j(\omega)$, and
permeabilities $\mu_j(\omega)$, as sketched in Fig.~\ref{multilayer}.
The $z$ axis is perpendicular to the layers, with the origin being on
the interface between layer $j$ $\!=$ $\!N-1$ and the free-space
region, which can be regarded as layer $j$ $\!=$ $\!N$ ($d_N$
$\!\to$ $\!\infty$, \mbox{$\varepsilon_N(\omega)$ $\!\equiv$ $\!1$},
\mbox{$\mu_N(\omega)$ $\!\equiv$ $\!1$)}. With the coordinate system
chosen such that the two atoms (in the free-space region) lie in the 
$xz$ plane, the nonzero elements of the scattering part
$\bm{G}^{(1)}(\mathbf{r}_A,\mathbf{r}_B,iu)$ of
the Green tensor $\bm{G}(\mathbf{r}_A,\mathbf{r}_B,iu)$ in
Eq.~(\ref{E31}) can be given by (App.~\ref{mulg})
\begin{multline}
\label{62}
\hspace{-2ex}{G}^{(1)}_{xx(yy)}({\mathbf r}_{A},
 {\mathbf r}_{B},iu)=\frac{1}{8\pi}\int_0^\infty {\rm d}q\,q
 e^{-bNZ_+}\\
\times\biggl[\frac{J_0(qX)\;\PM\,
 J_2(qX)}{b_N}\,r_N^s 
 -\frac{b_N[J_0(qX)
 \;\MP\,J_2(qX)]}{k_N^2}\,r_N^p\biggr],
\end{multline}
\begin{align}
\label{63}
&{G}^{(1)}_{xz(zx)}({\mathbf r}_{A},
 {\mathbf r}_{B},iu)=\MP
 \frac{1}{4\pi}\!\int_0^\infty\!\! {\rm d}q\,q^2
 e^{-b_NZ_+}\frac{J_1(qX)}{ k_N^2}\,r_N^{p},
\\
\label{64}
&{G}^{(1)}_{zz}({\mathbf r}_{A},{\mathbf r}_{B},iu)
 =-\frac{1}{4\pi}\int_0^\infty {\rm d}q\,q^3
 e^{-b_NZ_+}\frac{J_0(qX)}{b_Nk_N^2}\,r_N^p,
\end{align}
where  $Z_+$ $\!=$ $\!z_{A}$ $\!+$ $\!z_{B}$,
$X$ $\!=$ $\!x_{B}$ $\!-$ $\!x_{A}$, $J_\nu(x)$ denotes Bessel
functions,
 and
\begin{align}
\label{bl}
&b_j= b_{j}(q,u)
 =\sqrt{\frac{u^2}{c^2}\varepsilon_j(iu)
 \mu_j(iu)+q^2}\,
 ,\\
\label{kl}
&k_j=k_{j}(q,u)
 =\sqrt{\varepsilon_j(iu)\mu_j(iu)}\,
 \frac{u}{c}=n_j(iu)\,\frac{u}{c}\,.
\end{align}
The (generalized) reflection coefficients $r_j^\sigma$ with respect
to the left boundary of the $j$th layer ($j$ $\!=$ $\!1,2,3,\ldots,N$)
can be obtained from the recurrence relation
\begin{align}
\label{rsp}
r^\sigma_{j}&=r^\sigma_{j}(q,u)
 \nonumber\\
&=\frac{\left(\!\frac{\lambda^\sigma_{j- 1}}
 {b_{j-1}}-\frac{\lambda^\sigma_j}
 {b_j}\!\right)+
 \left(\!\frac{\lambda^\sigma_{j-1}}
 {b_{j-1}}+\frac{\lambda^\sigma_j}
 {b_j}\!\right)e^{-2b_{j- 1}d_{j-1}}
 r^\sigma_{j-1}}
 {\left(\!\frac{\lambda^\sigma_{j-1}}
 {b_{j-1}}+\frac{\lambda^\sigma_j}
 {b_j}\!\right)+\left(\!
 \frac{\lambda^\sigma_{j-1}}
 {b_{j-1}}-\frac{\lambda^\sigma_j}
 {b_j}\right)e^{-2b_{j- 1}d_{j-1}}
 r^\sigma_{j-1}}\,,
\end{align}
$r_{0}^\sigma$ $\!=$ $\!0$ ($\sigma$ $\!=$ $\!s,p$), where
$\lambda_j^{s}$ and $\lambda_j^p$ stand for $\mu_j(iu)$ and
$\varepsilon_j(iu)$, respectively.

According to the decomposition (\ref{gl64}) of the Green tensor, the
two-atom potential $U_{AB}$, Eq.~(\ref{E31}), can be decomposed into
three parts,
\begin{eqnarray}
\label{uparts}
U_{AB}(\mathbf{r}_A,\mathbf{r}_B)
&=&U_{AB}^{(0)}(\mathbf{r}_A,\mathbf{r}_B)
 +U_{AB}^{(1)}(\mathbf{r}_A,\mathbf{r}_B)\nonumber\\
&&+U_{AB}^{(2)}(\mathbf{r}_A,\mathbf{r}_B),
\end{eqnarray}
where
\begin{multline}
\label{gl72}
U_{AB}^{(0)}(\mathbf r_{A},\mathbf r_{B})=
 -\frac{\hbar\mu_0^2}{2\pi}\int_0^\infty {\mathrm  d}
 u\,u^4\alpha_{A}(iu)\alpha_{B}(iu)\\
\times\,
 \mathrm{Tr}\bigl[{\bm G}^{(0)}(\mathbf r_{A},
 \mathbf r_{B},iu)\!\cdot\!{\bm G}^{(0)}
 (\mathbf r_{B},
 \mathbf r_{A},iu)\bigr]
\end{multline}
is the bulk-part contribution, which is given by Eq.~(\ref{Bpot}) with
$n(iu)$ $\!\equiv$ $\!1$ $\!\equiv$ $\!\mu(iu)$,
\begin{multline}
\label{81}
U_{AB}^{(1)}(\mathbf r_{A},\mathbf
r_{B})=-\frac{\hbar\mu_0^2}{\pi}\int_0^\infty {\mathrm d}u\,
u^4\alpha_{A}(iu)\alpha_{B}(iu)\\
\times\,
\mathrm{Tr}\big[{\bm G}^{(0)}(\mathbf r_{A},
 \mathbf r_{B},iu)\!\cdot\!{\bm G}^{(1)}
 (\mathbf r_{B},\mathbf r_{A},iu)\big]\\
=-\frac{\hbar\mu_0^2}{32\pi^3 l}\int_0^\infty {\mathrm  d}
u\,u^4\alpha_{A}(iu)\alpha_{\mathrm
B}(iu)\,e^{-lu/c}\int_0^\infty {\rm d}q\,q \\
\times\,
e^{-b_NZ_+}
\left(
\bigg\{\bigg[2f(\xi)-g
(\xi)
\frac{X^2}{l^2}\bigg]\bigg[\frac{r_N^s}{b_N}-\frac{b_N}{k_N^2}\,r_N^p
\bigg]\right.\\
-2\bigg[f(\xi)-g(\xi)
\frac{Z^2}{l^2}\bigg]
\frac{q^2}{b_Nk_N^2}\,r_N^p\bigg\}J_0(qX)\\
\left.
-g(\xi)\frac{X^2}{l^2}\bigg[\frac{r_N^s}{b_N}+
\frac{b_N}{k_N^2}\,r_N^p\bigg]J_2(qX)
\right)
\end{multline}
comes from the cross term of bulk and scattering parts [with $f(x)$
and $g(x)$ being defined by Eqs.~(\ref{f}) and (\ref{g}), respectively,
$\xi$ $\!=$ $\!c/(lu)$, and $Z$ $\!=$ $\!z_{B}$ $\!-$ $\!z_{A}$], and
\begin{multline}
\label{82}
U_{AB}^{(2)}(\mathbf r_{A},\mathbf r_{B})
 =-\frac{\hbar\mu_0^2}{2\pi}\int_0^\infty \!{\mathrm d}u\,
 u^4\alpha_{A}(iu)\alpha_{B}(iu)\\
\times\,
 \mathrm{Tr}\big[{\bm G}^{(1)}(\mathbf r_{A},
 \mathbf r_{B},iu)\!\cdot\!{\bm G}^{(1)}
 (\mathbf r_{B},\mathbf r_{A},iu)\big]\\
=-\frac{\hbar\mu_0^2}{64\pi^3}\int_0^\infty {\mathrm  d}u\,
 u^4\alpha_{A}(iu)\alpha_{\mathrm B}(iu)
 \int_0^\infty \!{\rm d}q\, q
 \int_0^\infty \!{\rm d}q'\,q' \\
\times
 e^{-(b_N+b_N')Z_+}
 \biggl\{
 \bigg[\frac{r_N^sr_N^{s\prime}}{b_Nb_N'}
 + \frac{r_N^pr_N^{p\prime}}{k_N^4}\bigg(b_Nb_N'
 +\frac{2q^2q^{\prime 2}}{b_Nb_N'}\bigg) \\
-\frac{b_N'
 r_N^sr_N^{p\prime}}{b_N{k_N}^2}-\frac{b_N
 r_N^{s\prime} r_N^p}{b_N'{k_N}^2}\bigg]
 J_0(qX)J_0(q' X)\\
+\frac{4qq'
 r_N^pr_N^{p\prime}}{k_N^4}
 J_1(qX)J_1(q' X)
 + \bigg[\frac{r_N^sr_N^{s\prime}}{b_Nb_N'} \\
+\frac{b_Nb_N'r_N^pr_N^{p\prime}}{k_N^4}
 +\frac{b_N'r_N^sr_N^{p\prime}}{b_N{k_N}^2}+\frac{b_N
 r_N^{s\prime} r_N^p}{b_N'{k_N}^2}\bigg] \\
\times
 J_2(qX)J_2(q' X)\biggr\}
\end{multline}
is the scattering-part contribution [$b_N'$ $\!=$ $\!b_N(q',u)$,
$r^{\sigma\prime}_N$ $\!=r^{\sigma}_N(q',u)$].


\subsubsection{Perfectly reflecting plate}
\label{per}

\begin{figure}
\noindent
\begin{center}
\includegraphics[width=.9\linewidth]{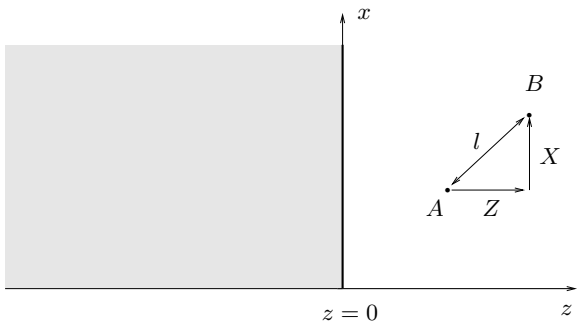}
\end{center}
\caption{
\label{reflector}
Two atoms near a perfectly reflecting plate.}
\end{figure}

Let us consider  the case $N$ $\!=$ $\!1$ (Fig.~\ref{reflector})
in more detail and begin with the limiting case of a perfectly
reflecting plate,
\begin{equation}
\label{rp-ref}
r_p\equiv r^p_{1}=\pm 1,\qquad r_s\equiv r^s_{1}=\mp 1,
\end{equation}
where the upper (lower) sign corresponds to a perfectly conducting
(permeable) plate. In the retarded limit, where $l,z_{\mathrm A},
z_{\mathrm B}$ $\!\gg$ $\!c/\omega_{\mathrm{min}}$
[$\omega_{\mathrm{min}}$ $\!=$
$\mathrm{min}(\{\omega_{A'}^n|A'\!=\!A,B;\,n\!=\!1,2,\ldots\})$],
$U_{AB}^{(0)}$ is given by Eq.~(\ref{E35}) with $n(0)$ $\!\equiv$
$\!1$ $\!\equiv$ $\!\mu(0)$, whereas $U_{AB}^{(1)}$ [Eq.~(\ref{81})]
and $U_{AB}^{(2)}$ [Eq.~(\ref{82})] can be given in closed form
only in some special cases. If $X$ $\!\ll$ $\!Z_+$
(cf.~Fig.~\ref{reflector}), we derive, on using the relevant elements of
the scattering Green tensor as given in App.~\ref{mulg}
[Eqs.~(\ref{rx}) and (\ref{rz})],
\begin{align}
\label{gl79}
& U_{AB}^{(1)}=
\pm\frac{32}{23}\,\frac{X^2+6l^2}{l^3Z_+(l+Z_+)^5}\,C_{\mathrm{r}}\,,
\\
\label{gl80}
& U_{AB}^{(2)}=-\frac{C_{\mathrm{r}}}{Z_+^7}\,,
\end{align}
where $C_{\mathrm{r}}$ is given by Eq.~(\ref{cr}) with
$\varepsilon(0)$ $\!\equiv$ $\!1$ $\!\equiv$ $n(0)$. Thus, recalling
Eq.~(\ref{E35}), the interaction potential (\ref{uparts}) reads
\begin{equation}
\label{gl801}
U_{AB}=
\Big[-\frac{1}{l^7}\pm\frac{32}{23}
\frac{X^2+6l^2}{l^3Z_+(l+Z_+)^5}-\frac{1}
{Z_+^7}\Big]C_{\mathrm{r}}\,.
\end{equation}
In particular, if $z_A$ $\!\ll$ $\! z_B$, or equivalently $Z_+$
$\!\simeq $ $\!Z\simeq l$, from Eqs.~(\ref{gl79}) and (\ref{gl80})
it follows that
\begin{align}
\label{gl81}
& U_{AB}^{(1)}=\mp\frac{6}{23}\,U_{AB}^{(0)}\,,
\\
\label{gl82}
& U_{AB}^{(2)}=U_{AB}^{(0)}\,,
\end{align}
so the interaction potential $U_{AB}$, Eq.~(\ref{uparts}), is
enhanced by the presence of the perfectly reflecting plate:
\begin{equation}
\label{gl83}
U_{AB}=
\left\{
\begin{array}{cc}
\frac{40}{23}U_{AB}^{(0)} & \ \mathrm{for}\ r_{p(s)}=
\PM 1,\\
&\\
\frac{52}{23}U_{AB}^{(0)} & \ \mathrm{for}\ r_{p(s)}=
\MP 1. \\
\end{array}
\right.
\end{equation}

Next, we discuss the behavior of $U_{AB}$ in the case where the condition
$z_A$ $\!\ll$ $\! z_B$ is not valid. Since the
bulk part $U_{AB}^{(0)}$ [first term in the square brackets in
Eq.~(\ref{gl801})] is negative, the interaction potential is enhanced
(reduced) by the plate if the scattering part
\mbox{$U_{AB}^{(1)}$ $\!+$ $\!U_{AB}^{(2)}$} [second and third terms
in the square brackets in Eq.~(\ref{gl801})] is negative (positive).
In the case of a perfectly conducting plate, it is seen that
especially for \mbox{$Z$ $\!=$ $\!0$}, briefly referred to as the
parallel case, \mbox{$U_{AB}^{(1)}$ $\!+$ $\!U_{AB}^{(2)}$} is
positive, and hence the interaction potential is reduced by the plate,
whereas for $X$ $\!=$ $\!0$, briefly referred to as the vertical case,
$U_{AB}^{(1)}$ $\!+$ $\!U_{AB}^{(2)}$ is positive and the interaction
potential is reduced iff
\begin{equation}
z_B/z_A\lesssim 4.90,
\end{equation}
where, without loss of generality, atom $A$ is assumed to be closer to
the plate than atom $B$. It is apparent from Eq.~(\ref{gl801}) that
for a perfectly permeable plate $U_{AB}^{(1)}$ $\!+$ $\!U_{AB}^{(2)}$
is always negative, and hence the interaction potential is always
enhanced by the plate.

Let us now turn to the nonretarded limit, where $l,z_{A},z_{B}$
$\!\ll$ $\!c/\omega_{\mathrm{max}}$ [$\omega_{\mathrm{max}}$ $\!=$
$\!\mathrm{max}(\{\omega_{A'}^n|A'\!=\!A,B;\,n$ $\!=1,2,\ldots\})$], and
$U_{AB}^{(0)}$ is given by Eq.~(\ref{E36}) [$\varepsilon(iu)$ $\!\equiv$
$\!1$]. From Eqs.~(\ref{81}) and (\ref{82}) we derive, 
on making use of the relevant elements of the scattering
Green tensor as given in App.~\ref{mulg}
[Eqs.~(\ref{xx})--(\ref{zz})],
\begin{align}
\label{e1}
&U_{AB}^{(1)}=\pm
\frac{4X^4-2Z^2Z_+^2+X^2(Z_+^2+Z^2)}{3l^5l_+^5}\,C_{\mathrm{nr}}\,,\\
\label{e2}
&U_{AB}^{(2)}=-\frac{C_{\mathrm{nr}}}{l_+^6}
\end{align}
($l_+$ $\!=$ $\!\sqrt{X^2+Z_+^2}$), where $C_{\mathrm{nr}}$ is given
by Eq.~(\ref{cnr}) with $\varepsilon(iu)$ $\!\equiv$ $\!1$. Hence,
the interaction potential (\ref{uparts}), reads, on
recalling Eq.~(\ref{E36}),
\begin{equation}
\label{full}
U_{AB} = \bigg[-\frac{1}{l^6}
\pm \frac{4X^4-2Z^2Z_+^2+X^2(Z_+^2+Z^2)}{3l^5l_+^5}
-\frac{1}{l_+^6}\bigg]C_{\mathrm{nr}}\,.
\end{equation}

Let us again consider the effect of the plate on the interaction
potential for the parallel and vertical cases. In the parallel case,
Eq.~(\ref{full}) takes the form
\begin{equation}
\label{par}
U_{AB}=\bigg[-\frac{1}{l^6}\pm
\frac{4l^2+Z_+^2}{3l^3(l^2+Z_+^2)^{\frac{5}{2}}}-
\frac{1}{(l^2+Z_+^2)^3}\bigg]C_{\mathrm{nr}}\,,
\end{equation}
which in the on-surface limit $Z_+$ $\!\to$ $\!0$ approaches
\begin{equation}
\label{gl95}
U_{AB}=
\left\{
\begin{array}{cc}
\frac{2}{3}U_{AB}^{(0)} & \ {\mathrm{for}}\ r_{p(s)}=\PM 1,\\
\\
\frac{10}{3}U_{AB}^{(0)} & \ {\mathrm{for}}\ r_{p(s)}=\MP 1. \\
\end{array}
\right.
\end{equation}
It can be seen easily that the term $U_{AB}^{(1)}$ [second term in the
square brackets in Eq.~(\ref{par})] dominates the term $U_{AB}^{(2)}$
[third term in the square brackets in Eq.~(\ref{par})], so
\mbox{$U_{AB}^{(1)}$ $\!+$ $\!U_{AB}^{(2)}$} is positive (negative)
for a perfectly conducting (permeable) plate, and hence the
interaction potential is reduced (enhanced) due to the
presence of the plate.

In the vertical case, from Eq.~(\ref{full}) the interaction potential
is obtained to be
\begin{equation}
\label{ver}
U_{AB}=\bigg[-\frac{1}{l^6}\mp\frac{2}{3Z_+^3l^3}
 -\frac{1}{Z_+^6}\bigg]C_{\mathrm{nr}}\,.
\end{equation}
It is obvious that \mbox{$U_{AB}^{(1)}$ $\!+$ $\!U_{AB}^{(2)}$}
[second and third terms in Eq.~(\ref{ver})] is negative when the plate
is perfectly conducting, thereby enhancing the interaction potential
since $U_{AB}^{(0)}$ [first term in Eq.~(\ref{ver})] is negative. In
the case of a perfectly permeable plate, $U_{AB}^{(1)}+U_{AB}^{(2)}$
is positive iff
\begin{equation}
\label{condition}
\frac{z_{B}}{z_{A}}<1+\frac{2}{(\frac{3}{2})^\frac{1}{3}-1}
 \simeq 14.82,
\end{equation}
where atom $A$ is again assumed to be closer to the plate than atom
$B$.

Since $U_{AB}^{(0)}$ and $U_{AB}^{(2)}$ are negative in all cases,
the realization of enhancement or reduction of the interaction
potential depends only on the sign of $U_{AB}^{(1)}$ and its magnitude
compared to that of $U_{AB}^{(2)}$.
\begin{figure}[t]
\noindent
\begin{center}
\includegraphics[width=.9\linewidth]{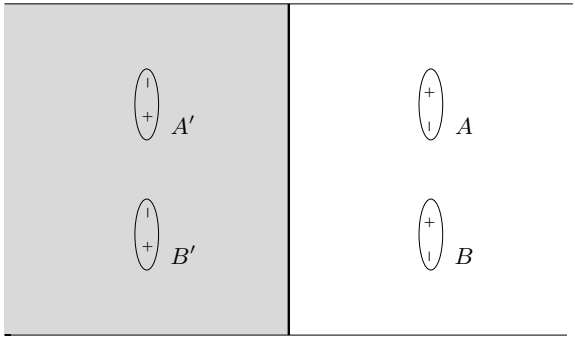}
\end{center}
\caption{
\label{con-par}
Two electric dipoles near a perfectly conducting plate (parallel
case).}
\end{figure}%
\begin{figure}[t]
\noindent
\begin{center}
\includegraphics[width=.9\linewidth]{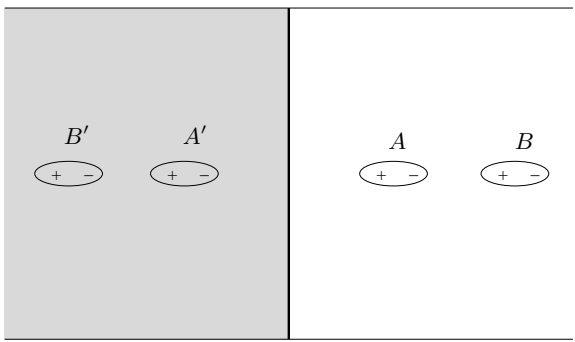}
\end{center}
\caption{
\label{con-ver}
Two electric dipoles near a perfectly conducting plate (vertical
case).}
\end{figure}%
In particular, the results for the non-retarded limit (the sign of
$U_{AB}^{(1)}$ being summarized in Tab.~\ref{sign}) can be explained
by using the method of image charges, where the two-atom vdW
interaction is regarded as being due to the interactions between
fluctuating dipoles $A$ and $B$ and their images ${A}'$ and ${B}'$
in the plate, with
\begin{equation}
 \hat H_{\mathrm {int}}=\hat V_{AB}+\hat V_{AB'}+\hat V_{BA'}
\end{equation}
being the corresponding interaction Hamiltonian. Here, $\hat{V}_{AB}$
denotes  the direct interaction between dipole $A$ and dipole $B$,
while $\hat V_{AB'}$ and $\hat V_{BA'}$ denote the indirect
interaction between each dipole and the image induced by the other one
in the plate. The leading contribution to the energy shift is of
second order in $H_{\mathrm {int}}$,
\begin{multline}
\label{deltaE}
\Delta E_{AB}=-\sideset{}{'}\sum_{n,m}
\frac{\langle 0_{A},0_{B}|\hat H_{\mathrm {int}}
|n_{A},m_{B}\rangle}{\hbar(\omega_{A}^n+
\omega_{B}^m)}
\\
\times\,\langle n_{A},
m_{B}|\hat H_{\mathrm {int}}|0_{A},
0_{B}\rangle.
\end{multline}

\begin{figure}[t]
\noindent
\begin{center}
\includegraphics[width=.9\linewidth]{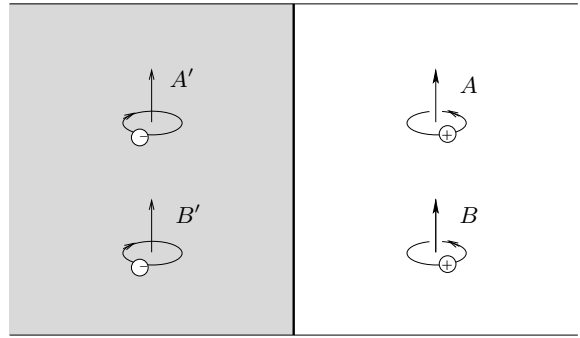}
\end{center}
\caption{
\label{mag-par}
Two magnetic dipoles near a perfectly conducting plate (parallel
case).}
\end{figure}

\begin{figure}[t]
\noindent
\begin{center}
\includegraphics[width=.9\linewidth]{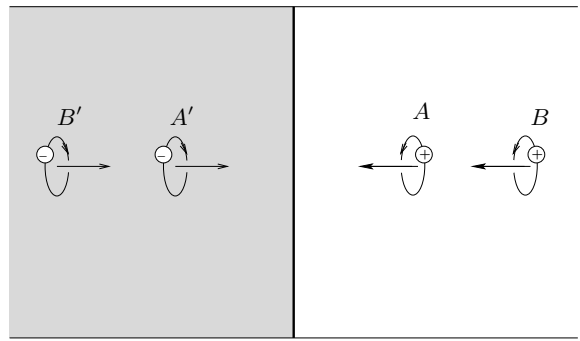}
\end{center}
\caption{
\label{mag-ver}
Two magnetic dipoles near a perfectly conducting plate (vertical
case).}
\end{figure}

In this approach, $U_{AB}^{(0)}$ corresponds to the product of two
direct interactions, so it is negative in agreement with
Eq.~(\ref{full}), because of the minus sign on the r.h.s.\ of
Eq.~(\ref{deltaE}). Accordingly, $U_{AB}^{(2)}$ is due to the product
of two indirect interactions and is also negative---in agreement with
Eq.~(\ref{full}). The terms containing one direct and one indirect
interaction are contained in $U^{(1)}_{AB}$ and determine its sign. We
can hence predict the sign of $U^{(1)}_{AB}$ from a graphical
construction of the image charges, as sketched in
Figs.~\ref{con-par}--\ref{mag-ver}.

Figure \ref{con-par} shows two electric dipoles in front of a
perfectly conducting plate in the parallel case. The configuration of
dipoles and images indicates repulsion between dipole $A(B)$ and
dipole $B'(A')$, so $U_{AB}^{(1)}$ is positive, in agreement with
Tab.~\ref{sign}. On the contrary, in the vertical case from
Fig.~\ref{con-ver} attraction is indicated, i.e., negative
$U_{AB}^{(1)}$, which is also in agreement with Tab.~\ref{sign}.
\begin{table}[b]
\begin{tabular}{|c|c|c|}
\hline
{} & conducting plate & permeable plate\\
\hline
parallel case & $+$ & $-$\\
\hline
vertical case & $-$ & $+$\\
\hline
\end{tabular}
\caption{
\label{sign}
Sign of $U_{AB}^{(1)}$ for a perfectly reflecting plate.}
\end{table}%

The case of two electric dipoles in front of a perfectly permeable
plate can be treated by considering two magnetic dipoles in front of a
perfectly conducting plate, as the two situations are equivalent due
to the duality between electric and magnetic fields in the absence of
free charges or currents. {F}rom Figs.~\ref{mag-par} (parallel case)
and \ref{mag-ver} (vertical case) it is apparent that the interaction
between dipole $A(B)$ and dipole $B'(A')$ is attractive in the
parallel case and repulsive in the vertical case, again confirming the
sign of $U_{AB}^{(1)}$ as given in Tab.~\ref{sign}. When the
dipole--dipole separation in Fig.~\ref{mag-ver} is sufficiently small
compared with the dipole--surface separations, then the direct
interaction between the two dipoles is expected to be stronger than
their indirect interaction via the image dipoles. As a result,
$U_{AB}^{(1)}$ will be the dominant term in \mbox{$U_{AB}^{(1)}$ $\!+$
$\!U_{AB}^{(2)}$} and $U_{AB}^{(1)}$ $\!+$ $\!U_{AB}^{(2)}$ becomes
positive. However, when the dipole--dipole separation exceeds the
dipole--surface separations, then the indirect interaction may become
comparable to the direct one, and $U_{AB}^{(2)}$ may be the dominant
term, leading to negative $U_{AB}^{(1)}+U_{AB}^{(2)}$. The image
dipole model hence gives also a qualitative explanation of the
condition~(\ref{condition}).


\subsubsection{Semi-infinite magnetodielectric half space}

Let us now abandon the assumption of perfect reflectivity and consider a
magnetodielectric plate of permittivity $\varepsilon(\omega)$ and
permeability $\mu(\omega)$. To be more specific, we restrict our
attention to a sufficiently thick plate so that the model of a
semi-infinite half space applies. In this case, Eq.~(\ref{rsp}) for the
reflection coefficients reduces to
\begin{equation}
\label{eq95}
r_\sigma\equiv r_{1}^\sigma=\frac{\lambda_0
^\sigma
b - b_0}{\lambda_0
^\sigma
 b+b_0}\,,
\end{equation}
with $b$ $\!\equiv$ $\!b_1$ $\!=$ $\!\sqrt{u^2/c^2\!+\!q^2}$, $b_0$
$\!=$ $\!\sqrt{\varepsilon(iu)\mu(iu)u^2/c^2\!+\!q^2}$,
\mbox{$\lambda_0^s$ $\!=$ $\!\mu(iu)$}, and $\lambda_0^p$ $\!=$
$\!\varepsilon(iu)$.

In the retarded limit, $l,z_A,z_B$ $\!\gg$ $\!c/\omega_{\mathrm
{min}}$ [with $\omega_\mathrm{min}$ being defined as above
Eq.~(\ref{alpha0})] we may again replace the atomic polarizability and
the permittivity and permeability of the plate by their static values.
Replacing the integration variable $q$ in Eq.~(\ref{81}) by $v$ $\!=$
$\!b_1c/u$ [cf.~Eq.~(\ref{C8})] leads to
\begin{align}
&U_{AB}^{(1)}(\mathbf{r}_A,\mathbf{r}_B)
=\frac{\hbar c}{32\pi^3 l^3\varepsilon_0^2}
\alpha_{A}(0)\alpha_{B}(0)\int_1^\infty
{\rm d}v\,
\nonumber\\&\quad\times
\bigg\{\bigg[v^2\bigg(Z^2A_{5-}+(Z^2-2X^2)\bigg[\frac{A_{4-}}{l}
+\frac{A_{3-}}{l^2}\bigg]
\nonumber\\&\qquad
+l^2A_{5+}+lA_{4+}+A_{3+}\bigg)+2(v^2-1)\bigg(X^2B_{5}
\nonumber\\&\qquad
+\big(X^2-2Z^2\big)\bigg[\frac{B_4}{l}+\frac{B_3}{l^2}\bigg]\bigg)
\bigg]r_{p}
\nonumber\\
&\qquad
+\bigg(Z^2A_{5+}+\big(Z^2-2X^2\big)\bigg[\frac{A_{4+}}{l}
+\frac{A_{3+}}{l^2}\bigg]
\nonumber\\&\qquad
+l^2A_{5-}+lA_{4-}+A_{3-}\bigg)r_{s}
\bigg\},
\end{align}
where, according to Eq.~(\ref{eq95}), the static reflection
coefficients are given by
\begin{align}
\label{static1}
&r_s=r_s(v)
=\frac{\mu(0)v-\sqrt{\varepsilon(0)\mu(0)-1+v^2}}
{\mu(0)v+\sqrt{\varepsilon(0)\mu(0)-1+v^2}}\,, \\
\label{static2}
&r_p=r_p(v)
=\frac{\varepsilon(0)v-\sqrt{\varepsilon(0)\mu(0)-1+v^2}}
{\varepsilon(0)v+\sqrt{\varepsilon(0)\mu(0)-1+v^2}}\,,
\end{align}
and
\begin{align}
\label{A}
&A_{n\pm}=
\frac{1}{c^{n+1}}\int_0^\infty\mathrm d
 u\,u^n\,e^{-au/c}\big[J_0(\beta u/c)\pm J_2(\beta u/c)\big],\\
\label{B}
&B_n=\frac{1}{c^{n+1}}\int_0^\infty{\mathrm d}u\,u^n
 e^{-au/c}J_0(\beta u/c),
\end{align}
with $\beta$ $\!=$ $\!X\sqrt{v^2-1}$ and  $a$ $\!=$ $\!l$ $\!+$ $\!vZ_+$
(for explicit expressions of $A_{n\pm}$ and $B_n$, see App.~\ref{an}).
Similarly, Eq.~(\ref{82}) reduces to
\begin{align}
\label{24}
&U_{AB}^{(2)}=-\frac{\hbar \mu_0^2}{64\pi^3 c^2}\,\alpha_{A}(0)
\alpha_{B}(0)\int_1^\infty {\rm d}v\int_1^\infty {\rm d}v'
\nonumber\\&\quad\times
\biggl\{
\Big(r_{p}r_p'
\big[3v^2v'^2-2(v^2+v'^2)+2\big]
+r_{s}r_s'-r_{s}r_p'v'^2
\nonumber\\&\quad
-r_pr_s'v^2\Big)M_0
+4vv'\sqrt{v^2-1}\sqrt{v'^2-1}r_{p}r_p'M_1
\nonumber\\&\quad
+\big(r_{s}r_s'+r_{p}r_p'v^2v'^2+r_{s}r_p'v'^2
+r_pr_s'v^2\big)M_2\biggr\}
\end{align}
[$r'_{\sigma}=r_{\sigma}(v')$], 
where
\begin{equation}
\label{M}
M_n=\int_0^\infty{\mathrm d}u\,u^6e^{-(v+v')Z_+ u/c}J_n(\beta
u/c)J_n(\beta'u/c)
\end{equation}
($\beta'$ $\!=$ $\!X\sqrt{v'^2-1}$), which can be evaluated
analytically only in some special cases. In particular, when $X$
$\!\ll$ $\! Z_+$, then approximately
\begin{equation}
M_n=
J_n^2(0)\int_0^\infty{\mathrm d}u\,
u^6e^{-(v+v')Z_+u/c}
= \frac{720c^7}{(v+v')^7Z_+^7}\,\delta_{n0}.
\end{equation}

In the nonretarded limit, $l,z_{A},z_{B}$ $\!\ll$
$\!c/[n(0)\omega_\mathrm{max}]$ [with $\omega_\mathrm{max}$ being 
defined as
above Eq.~(\ref{E36})], $U_{AB}^{(1)}$ and $U_{AB}^{(2)}$ can be
obtained by using in Eqs.~(\ref{81}) and (\ref{82}), respectively, the
relevant elements of the scattering part of Green tensor as given in
App.~\ref{mulg}. In the case of a purely dielectric half space ($\mu$
$\!\equiv$ $\!1$) we derive [Eqs.~(\ref{Axx})--(\ref{Bzz})]
\begin{multline}
\label{udie}
\hspace{-2ex}
U_{AB}=-\frac{C_{\mathrm{nr}}}{l^6}
\\
+\frac{\big[4X^4-2Z^2Z_+^2+
X^2(Z^2+Z_+^2)\big]
C^{(1)}_\mathrm{nr}}{l^5l_+^5}-\frac{C^{(2)}_\mathrm{nr}}{l_+^6}\,,
\end{multline}
where $C_{\mathrm{nr}}$ is given by Eq.~(\ref{cnr})
with $\varepsilon(iu)$ $\!\equiv$ $\!1$, and
\begin{align}
\label{c1}
&\hspace{-1ex}C^{(1)}_\mathrm{nr}=\frac{\hbar}{16\pi^3\varepsilon_0^2}
\int_0^\infty{\mathrm d}u\,\alpha_{A}(iu)
\alpha_{\mathrm
B}(iu)\frac{\varepsilon(iu)-1}{\varepsilon(iu)+1}\,,
\\
\label{c2}
&\hspace{-1ex}C^{(2)}_\mathrm{nr}
=\frac{3\hbar}{16\pi^3\varepsilon_0^2}\int_0^\infty{\rm
d}u\,\alpha_{A}(iu)\alpha_{B}(iu)
\bigg[\frac{\varepsilon(iu)-1}{\varepsilon(iu)+1}
\bigg]^2\!.
\end{align}
In particular in the limiting case when $l$ $\!\ll$ $\!Z_+$, 
Eq.~(\ref{udie}) reduces to
\begin{equation}
\label{udie2}
U_{AB}=-\frac{C_{\mathrm{nr}}}{l^6}
+\frac{\big(X^2-2Z^2\big)
C^{(1)}_\mathrm{nr}}{l^5Z_+^3}\,.
\end{equation}
It is seen that the second term on the r.h.s.\ of this equation is
positive  (negative) in the parallel (vertical) case, so the vdW
potential is reduced (enhanced) by the presence of the dielectric half
space. In the case of a purely magnetic half space ($\varepsilon$
$\!\equiv$ $\!1$) we derive [Eqs.~(\ref{Axx2})--(\ref{Azz})]
\begin{equation}
\label{umag1}
U_{AB}=
-\frac{C_\mathrm{nr}}{l^6} +
\frac{\big[Z^2-2X^2+3Z_+(l_+-Z_+)\big]
 C_\mathrm{nr}^{(3)}}{l^5l_+}\,,
\end{equation}
where
\begin{multline}
C_\mathrm{nr}^{(3)}
=\frac{\hbar}{64\pi^3\varepsilon_0^2c^2}\int_0^\infty
{\rm d}u\,u^2\alpha_{A}(iu)\alpha_{B}(iu)\\
\times\;\frac{[\mu(iu)-1][\mu(iu)-3]}{\mu(iu)+1}\,.
\end{multline}
Note that $U_{AB}^{(2)}$ does not contribute to the asymptotic
nonretarded two-atom vdW potential $U_{AB}$ for the purely magnetic
half space. In particular in the limiting case when $X$ $\!\ll$
$\!Z_+$, Eq.~(\ref{umag1}) reduces to
\begin{equation}
\label{umag2}
U_{AB}=
-\frac{C_\mathrm{nr}}{l^6} +
\frac{\big(2Z^2-X^2\big)C_\mathrm{nr}^{(3)}}{2l^5Z_+}\,.
\end{equation}
It is seen that the second term in the r.h.s.\ of this equation is
negative (positive) in the  parallel (vertical) case, so the vdW
potential is enhanced (reduced) due to the presence of the magnetic
half space.

It should be pointed out that the nonretarded limit for the
magnetodielectric half space is in general incompatible with the limit
of perfect reflectivity [\mbox{$\varepsilon(iu)\rightarrow \infty$} or
\mbox{$\mu(iu)$ $\!\to$ $\!\infty$}] considered in Sec.~\ref{per}, as
is clearly seen from the condition given above Eq.~(\ref{udie})
[cf.\ also the expansions (\ref{rs-lim}) and (\ref{rp-lim}), which
are not well-behaved in the limit of perfect reflectivity]. As a
consequence, Eq.~(\ref{umag1}) does not reduce to Eq.~(\ref{full}) via
the limit \mbox{$\mu(iu)$ $\!\to$ $\!\infty$}. It is therefore
remarkable that the result for a purely dielectric half space,
Eq.~(\ref{udie}), does reduce to Eq.~(\ref{full}) in the limit
\mbox{$\varepsilon(iu)$ $\!\to$ $\!\infty$}, as already noted in
Ref.~\cite{Babiker76} in the case of the single-atom potential.

\begin{figure}
\noindent
\begin{center}
\includegraphics[width=\linewidth]{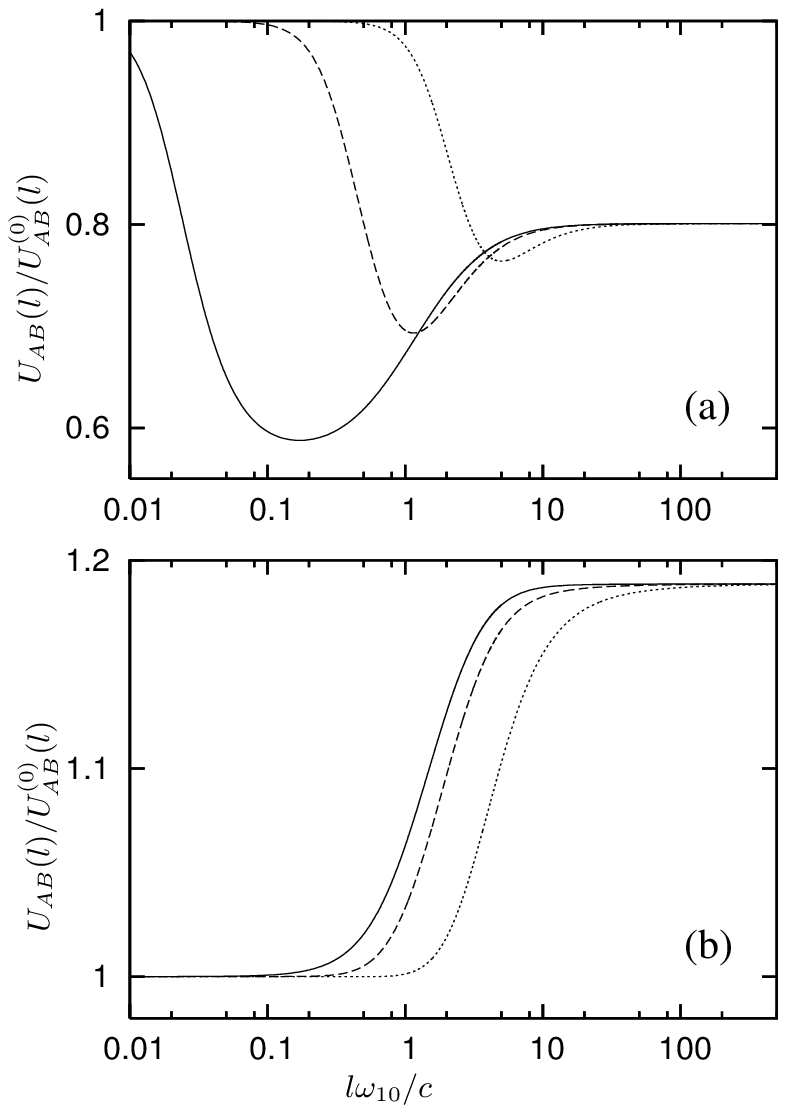}
\end{center}
\caption{
\label{p-par}
The vdW potential for two identical two-level atoms in the parallel
case in presence of (a) a  purely
dielectric half space with
\mbox{$\omega_{\mathrm{P}e}/\omega_{10}$ $\!=$ $\!3$},
\mbox{$\omega_{\mathrm{T}e}/\omega_{10}$ $\!=$ 
$\!1$},
and
\mbox{$\gamma_e/\omega_{10}$ $\!=$ $\!0.001$}
(b) a purely magnetic half space with
\mbox{$\omega_{\mathrm{P}m}/\omega_{10}$ $\!=$ $\!3$},
\mbox{$\omega_{\mathrm{T}m}/\omega_{10}$ $\!=$ $\!1$},
and \mbox{$\gamma_m/\omega_{10}=0.001$} is shown as a function of the
atom--atom separation $l$ [$\omega_{10}$ is the atomic
transition frequency, and $U_{AB}^{(0)}(l)$ is the potential in free
space]. The atom--half-space separations are $z_{A}\!=\!z_{B}$
$\!=$ $\!0.01c/\omega_{10}$ (solid line), $0.2c/\omega_{10}$ (dashed
line), and $c/\omega_{10}$ (dotted line).
}
\end{figure}%

\begin{figure}
\noindent
\begin{center}
\includegraphics[width=\linewidth]{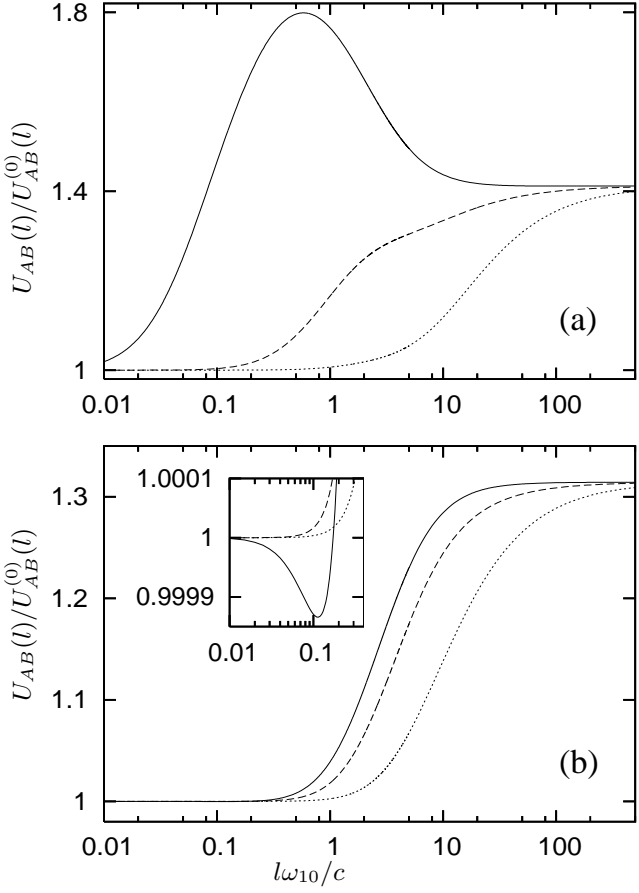}
\end{center}
\caption{
\label{p-ver}
The vdW potential for two two-level atoms in the vertical case
in the presence of (a) a purely dielectric half space and (b) a purely
magnetic half space is shown as a function of the atom--atom
separation $l$. The distance between atom $A$ (which is closer to the
surface of the half space than atom $B$) and the surface is equal to
$0.01c/\omega_{10}$ (solid line), $0.2c/\omega_{10}$ (dashed line),
and $c/\omega_{10}$ (dotted line). All other parameters are the same
as in Fig.~\ref{p-par}.}
\end{figure}

\begin{figure}
\noindent
\begin{center}
\includegraphics[width=\linewidth]{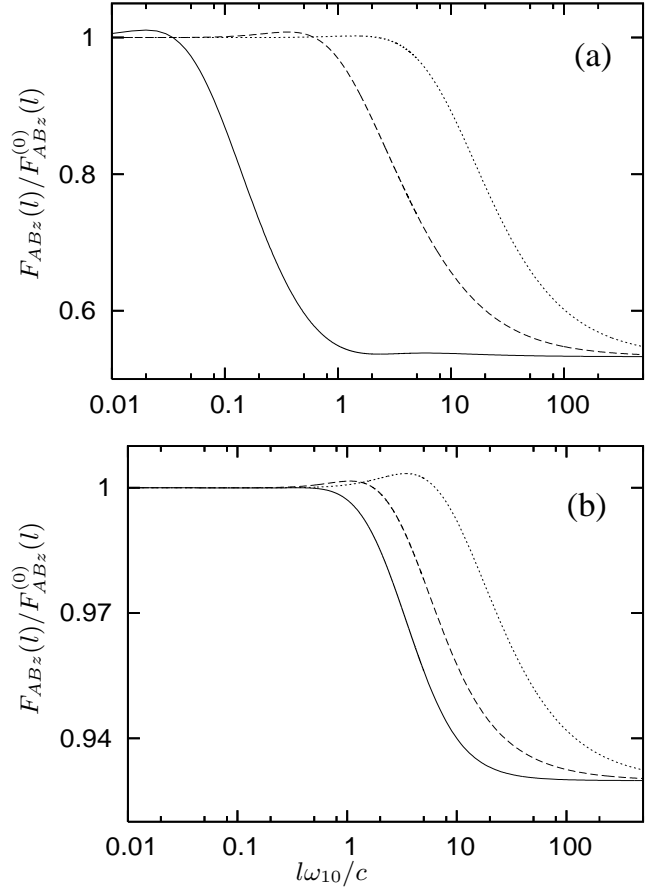}
\end{center}
\caption{
\label{f-a-ver}
The vdW force acting on atom $A$ (which is closer to the surface of
the half space than atom $B$) in the presence of (a) a purely
dielectric half space and (b) a purely magnetic half space is shown as
a function of the atom--atom separation $l$. All parameters are the
same as in Fig.~\ref{p-ver}.}
\end{figure}

Figures~\ref{p-par}--\ref{f-a-ver} show the results of exact
(numerical) calculation of the vdW interaction between two identical
two-level atoms near a semi-infinite half space, as given by
Eq.~(\ref{uparts}) together  with Eqs.~(\ref{Bpot}), (\ref{81}), and
(\ref{82}). In the figures the potentials and the forces are
normalized w.r.t.\ their values in free space as given by
Eq.~(\ref{Bpot}) \mbox{[$n(iu)$ $\equiv$ $\!1$ $\!\equiv$
$\!\mu(iu)$]}. In the calculations, we have used single-resonance
Drude--Lorenz-type electric and magnetic susceptibilities of the half
space,
\begin{equation}
\label{e109}
\varepsilon(\omega)=1+\frac{\omega_{\mathrm{P}e}^2}{\omega^2_{{\mathrm
{T}e}}
-\omega^2-i\omega\gamma_e}\,,
\end{equation}
\begin{equation}
\label{e110}
\mu(\omega)=1+\frac{\omega_{\mathrm{P}m}^2}{\omega^2_{\mathrm{T}m}
-\omega^2-i\omega\gamma_m}\,.
\end{equation}
From the figures it is seen that the vdW interaction is unaffected by
the presence of the half space for atom--half-space separations that
are much greater than the interatomic separations, while an asymptotic
enhancement or reduction of the interaction is observed in the
opposite limit.

Figure~\ref{p-par}(a) shows the dependence of the normalized
interaction potential $U_{AB}(l)$ on the atom--atom separation $l$ in
the parallel case \mbox{($Z$ $\!=$ $\!0$)} for different values of the
distance \mbox{$z_A$ $\!(=$ $\!z_B)$} of the atoms from a purely
dielectric half space. The ratio of the interatomic force along the
connecting line of the two atoms, $F_{ABx}(l)$ [Eq.~(\ref{inforce})]
to the corresponding force in free space, $F^{(0)}_{ABx}(l)$, follows
closely the ratio $U_{AB}(l)/U^{(0)}_{AB}(l)$, so that, within the
resolution of the figures, the curves for
$F_{ABx}(l)/F^{(0)}_{ABx}(l)$ (not shown) would  coincide with those
for $U_{AB}(l)/U^{(0)}_{AB}(l)$. The figure reveals that due to the
presence of the dielectric half space the attractive interaction
potential and force are reduced, in agreement with the predictions
from the nonretarded limit, Eq.~(\ref{udie2}). The relative reduction
of the potential and the force are not monotonic, there is a value of
the atom--atom separation where the reduction is strongest. The
$l$-dependence  of $U_{AB}(l)/U^{(0)}_{AB}(l)$ in the presence of a
purely magnetic half space in the parallel case is shown in
Figs.~\ref{p-par}(b). Again, the corresponding force ratio
$F_{ABx}(l)/F^{(0)}_{ABx}(l)$ (not shown) behaves like
$U_{AB}(l)/U^{(0)}_{AB}(l)$. The figure indicates
that the presence of a purely magnetic half space enhances the vdW
interaction between the two atoms, with the enhancement increasing
with the atom-atom separation, in agreement with the nonretarded
limit, Eq.~(\ref{umag2}).

Figure~\ref{p-ver} shows $U_{AB}(l)/U^{(0)}_{AB}(l)$ in the vertical
case \mbox{($X$ $\!=$ $\!0$)} when the half space is purely dielectric
[Fig.~\ref{p-ver}(a)] or purely magnetic [Fig.~\ref{p-ver}(b)]. In the
figure, atom $A$ is assumed to be closer to the surface of the half
space than atom $B$, and the graphs show the variation of the
interaction potential with the atom--atom separation $l$ for different
distances $z_A$ of atom $A$ from the surface of the half space. It is
seen that for a purely dielectric half space the potential is enhanced
compared to the one observed in the free-space case---in agreement 
with Eq.~(\ref{udie2}). Note that there are values of the
atom--atom separation at which the enhancement is strongest.
For a purely magnetic half space, the potential is seen to be
typically enhanced although for very small atom--atom separations a
reduction appears [inset in Fig.~\ref{p-ver}(b)]---in agreement with
Eq.~(\ref{umag2}). As in the parallel case, the relative enhancement
is monotonic. Whereas the force $F_{BAz}(l)/F^{(0)}_{BAz}(l)$ for the
force acting on atom $B$ (not shown) again follows closely the
potential ratio $U_{AB}(l)/U^{(0)}_{AB}(l)$, the ratio
$F_{ABz}(l)/F^{(0)}_{ABz}(l)$, for the force acting on atom $A$
noticeably differs from $U_{AB}(l)/U^{(0)}_{AB}(l)$, as can be seen
from comparing Figs.~\ref{p-ver} and \ref{f-a-ver}. Clearly, the
reason must be seen in the different atom--atom and atom--half-space
directions in the two cases (cf.~Figs.~\ref{con-ver} and
\ref{mag-ver}).

Figs.~\ref{p-par}(a) and \ref{p-ver}(a) showing the interaction
potential of two atoms in the presence of a purely dielectric half
space in the parallel and vertical cases, respectively cover the
results shown in Ref.~\cite{Cho} on a different scale. The results
here are more complete because they show that the relative potential
does not have the monotonic behavior suggested by the figures in
Ref.~\cite{Cho}.


\section{Summary and Conclusions}
\label{con}

Based on macroscopic QED in linear, causal media, we have obtained a
general formula for the vdW potential of two ground-state atoms
in the presence of an arbitrary arrangement of dispersing and
absorbing magnetodielectric media by calculating the leading-(4th)
order shift of the ground-state energy of the overall system.
The result has been applied to two atoms (i) in bulk material
(without taking into account local-field corrections), (ii) in the
presence of a perfectly reflecting plate, and (iii) in the presence of
a semi-infinite magnetodielectric half space. It has been found that
the presence of a bulk magnetodielectric medium will reduce the
interaction potential w.r.t. its well-known free-space value.

We have further shown that in the presence of a perfectly reflecting
plate the vdW interaction can be enhanced or reduced depending on the
(electric/magnetic) nature of the plate and the (parallel/vertical)
alignment of the atoms. In particular, in the nonretarded limit these
effects can be qualitatively explained using the method of image
dipoles.

Finally, we have calculated the vdW potential in the presence of a
magnetodielectric half space. The analytical results show that in the
nonretarded limit the potential in the case of a purely
dielectric half space is reduced (enhanced) in the parallel (vertical)
case compared to its value in free space, while in the case of a
purely magnetic half space it is enhanced (reduced) for parallel
(vertical) alignment of the two atoms. The results for a purely
dielectric half space are in qualitative agreement with those for the
perfectly conducting plate, while for a magnetic plate the results for
finite permeability disagree with those for the perfectly reflecting
case in the asymptotic power laws---owing to the fact that the two
limits of perfect reflectivity and nonretarded distance do not
commute.

The numerical computation of the interaction potential in the whole
distance regime confirms the analytical results. In addition, it
shows that the relative enhancement/reduction of the vdW interaction
is not always monotonous, but may in general display maxima or minima,
in particular in the case of a purely dielectric half space.


\begin{acknowledgments}
This work was supported by the Deutsche Forschungsgemeinschaft.
H.S. would like to thank the Ministry of Science, Research, and
Technology of Iran. H.T.D. thanks T. Kampf for a helpful hint on
programming. He would also like to thank the Alexander von
Humboldt Stiftung and the National Program for Basic Research of
Vietnam.
\end{acknowledgments}


\begin{appendix}


\section{Intermediate states and interaction matrix elements}
\label{mat-elm}

The intermediate states contributing to the two-atom vdW interaction
according to Eq.~(\ref{E12}) are listed in the first three columns of
Tab.~\ref{denom1}; the corresponding matrix elements of the
interaction Hamiltonian~(\ref{eq49-1}) [together with Eqs.~(\ref{eq8})
and (\ref{eq20})] can be found by recalling the commutation relations
(\ref{eq3.1}) and (\ref{eq3.2}) and using the relations (\ref{Gprop2})
and (\ref{Gprop3}). For example, for case~(1) in Tab.~\ref{denom1}
this leads to

\begin{widetext}
\begin{center}
\begin{table}[ht]
\begin{tabular}{cllll}
\hline
 Case  & $|I\rangle$    &
 \hspace{1ex}
 $|II\rangle$   &
 \hspace{1ex}
 $\hspace{-1ex}
 |III\rangle$ &
 Denominator\\
\hline
($1$)
& $|n_A,0_B\rangle |1^{(1)}\rangle$
      &\hspace{1ex} $|0_A,0_B\rangle
        |1^{(2)},1^{(3)}\rangle$
      & \hspace{1ex}$|0_A,m_B\rangle |1^{(4)}\rangle$
      & $D_{\mathrm {i}}=(\omega_A^n+\omega')
      (\omega'+\omega)(\omega_B^m+\omega')$,  \\
      {}
      & ${}$
      & ${}$
      & ${}$
      & $D_{\mathrm {ii}}=(\omega_A^n+\omega')
      (\omega'+\omega)(\omega_B^m+\omega)$  \\
($2$)
      &-''-
      & \hspace{2ex}$|n_A,m_B\rangle |\{0\}\rangle$
      & \hspace{1ex}$|0_A,m_B\rangle |1^{(2)}\rangle$
      & $D_{\mathrm {iii}}=(\omega_A^n+\omega')
        (\omega_A^n+\omega_B^m)
        (\omega_B^m+\omega)$\\
($3$)
      & -''-
      & \hspace{2ex}-''-
      &\hspace{1ex}$|n_A,0_B\rangle |1^{(2)}\rangle$
      & $D_{\mathrm {iv}}=(\omega_A^n+\omega')
         (\omega_A^n+\omega_B^m)
            (\omega_A^n+\omega)$\\
($4$)
      & -''-
      & \hspace{2ex}$|n_A,m_B\rangle
        |1^{(2)},1^{(3)}\rangle$
      & \hspace{1ex}$|0_A,m_B\rangle |1^{(4)}\rangle$
      & $D_{\mathrm {v}}=(\omega_A^n+\omega')
         (\omega_A^n+\omega_B^m+\omega'+
         \omega)
     (\omega_B^m+\omega')$\\
($5$)
      &-''-
      & \hspace{2ex}-''-
      & \hspace{1ex}$|n_A,0_B\rangle |1^{(4)}\rangle$
      & $D_{\mathrm {vi}}=(\omega_A^n+\omega')
         (\omega_A^n+\omega_B^m+
         \omega'+\omega)
     (\omega_A^n+\omega)$\\
($6$)
      & $|0_A,m_B\rangle |1^{(1)}\rangle$
      & \hspace{2ex}$|0_A,0_B\rangle
        |1^{(2)},1^{(3)}\rangle$
      & \hspace{1ex}$|n_A,0_B\rangle |1^{(4)}\rangle$
            & $D_{\mathrm {vii}}=(\omega_B^m+\omega')
      (\omega'+\omega)(\omega_A^n+\omega')$,  \\
{}
      & ${}$
      & ${}$
      & ${}$
           & $D_{\mathrm{viii}}=(\omega_B^m+\omega')
      (\omega'+\omega)(\omega_A^n+\omega)$  \\
($7$)
      & -''-
      &\hspace{1ex} $|n_A,m_B\rangle |\{0\}\rangle$
      & \hspace{1ex}$|n_A,0_B\rangle |1^{(2)}\rangle$
      & $D_{\mathrm {ix}}=(\omega_B^m+\omega')
        (\omega_A^n+\omega_B^m)
        (\omega_A^n+\omega)$\\
($8$)
      & -''-
      & \hspace{2ex}-''-
      & \hspace{1ex}$|0_A,m_B\rangle |1^{(2)}\rangle$
        & $D_{\mathrm {x}}=(\omega_B^m+\omega')
         (\omega_A^n+\omega_B^m)
            (\omega_B^m+\omega)$\\
($9$)
      & -''-
      & \hspace{2ex}$|n_A,m_B\rangle
        |1^{(2)},1^{(3)}\rangle$
      & \hspace{1ex}$|n_A,0_B\rangle |1^{(4)}\rangle$
      & $D_{\mathrm {xi}}=(\omega_B^m+\omega')
         (\omega_A^n+\omega_B^m+\omega'+\omega)
     (\omega_A^n+\omega')$\\
($10$)
      &-''-
      & \hspace{2ex}-''-
      & \hspace{1ex}$|0_A,m_B\rangle |1^{(4)}\rangle$
      & $D_{\mathrm {xii}}=(\omega_B^m+\omega')
         (\omega_A^n+\omega_B^m+\omega'+
         \omega)
     (\omega_B^m+\omega)$\\
\hline
\end{tabular}
\caption{
\label{denom1}
Intermediate states contributing to the two-atom vdW potential and
corresponding denominators.
}
\end{table}
\end{center}
\end{widetext}%

\vspace*{-10ex}
\begin{align}
\label{mel1}
&\langle 1^{(1)}|\langle n_{A}|\langle 0_{B}|
 H_{A\mathrm{F}}+H_{B\mathrm {F}}
 |0_{A}\rangle| 0_{B}\rangle|\{0\}\rangle\nonumber\\
&\quad=-\big[\mathbf{d}^{n0}_A\!\cdot\!\bm{G}^\ast_{\lambda_1}
 (\mathbf{r}_{A},\mathbf{r}_1,\omega_1)\big]_{i_1},
\\[1ex]
\label{mel2}
&\langle 1^{(2)},1^{(3)}|\langle 0_{A}|\langle 0_{B}|
 H_{A\mathrm{F}}+H_{B\mathrm{F}}
 |n_{A}\rangle|0_{B}\rangle| 1^{(1)}\rangle\nonumber\\
&\quad=-\frac{1}{\sqrt{2}}\Big\{\big[\mathbf{d}_A^{0n}\!\cdot\!
 \bm{G}^\ast_{\lambda_3}(\mathbf{r}_{A},\mathbf{r}_3,\omega_3)
 \big]_{i_3}\delta^{(12)}\nonumber\\
&\qquad\,+\big[\mathbf{d}_A^{0n}\!\cdot\!\bm{G}^\ast_{\lambda_2}
 (\mathbf{r}_{A},\mathbf{r}_2,\omega_2)
 \big]_{i_2}\delta^{(13)}\Big\},
\end{align}
\begin{align}
\label{mel3}
&\langle 1^{(4)}|\langle 0_{A}|\langle m_{B}|
 H_{A\mathrm{F}}+H_{B\mathrm{F}}
 |0_{A}\rangle|0_{B}\rangle|1^{(2)},1^{(3)}\rangle\nonumber\\
&\quad =-\frac{1}{\sqrt{2}}\Big\{\big[\mathbf d_B^{m0}\!\cdot\!
\bm G_{\lambda_3}(\mathbf r_{B},\mathbf r_3,\omega_3)
\big]_{i_3}\delta^{(24)}\nonumber\\
&\qquad\,+\big[\mathbf d_B^{m0}\!\cdot\!
\bm G_{\lambda_2}(\mathbf r_{B},\mathbf r_2,\omega_2)
\big]_{i_2}\delta^{(34)}\Big\},
\end{align}
\begin{align}
\label{mel4}
&\langle\{0\}|\langle 0_{A}|\langle 0_{B}|
 H_{A\mathrm{F}}+H_{B\mathrm {F}}
 |0_{A}\rangle |m_B\rangle|1^{(4)}\rangle\nonumber\\
&\quad =-\big[\mathbf d_B^{0m}\!\cdot\!
 \bm G_{\lambda_4}
(\mathbf r_{B},\mathbf r_4,\omega_4)\big]_{i_4},
\end{align}
where $\delta^{(\alpha\beta)}$ is given by Eq.~(\ref{delta}).
Substituting them into Eq.~(\ref{E12}), one obtains Eq.~(\ref{wow})
and subsequently Eq.~(\ref{E20}), with energy denominators
$D_{\mathrm i}$ and $D_{\mathrm {ii}}$ as given in Tab.~\ref{denom1}.
The other denominators listed in the last column of the table follow
in a similar way from the respective intermediate states
given in the first three columns.


\section{Derivation of Eq.~(\ref{E21})}
\label{AppA}

From the energy denominators given in Tab.~\ref{denom1}, it is
straightforward to obtain
\begin{multline}
\label{A1}
\frac{1}{D_\mathrm{ii}}+\frac{1}{D_\mathrm{iii}}
 +\frac{1}{D_\mathrm{viii}}+\frac{1}{D_\mathrm{ix}}
 +\frac{1}{D_\mathrm{iv}}+\frac{1}{D_\mathrm{x}}
\\
=\frac{1}{\omega_A^n\!+\!\omega_B^m}
 \Biggl[\biggl(\frac{1}{\omega_A^n\!+\!\omega} +
 \frac{1}{\omega_B^m\!+\!\omega} \biggr)
 \biggl(\frac{1}{\omega\!+\!\omega'}
 -\frac{1}{\omega\!-\!\omega'}\biggr)
\\
\quad +\biggl(\frac{1}{\omega_A^n\!+\!\omega'}
 +\frac{1}{\omega_B^m\!+\!\omega'} \biggr)
 \biggl(\frac{1}{\omega\!+\!\omega'}
 + \frac{1}{\omega\!-\!\omega'}\biggr)
 \Biggr].
\end{multline}
Since the denominators appear in combinations of the form of
Eq.~(\ref{E20}), where they are multiplied with terms (the two
factors in square brackets) which are always the same and symmetric
with respect to $\omega$ and $\omega'$, we may interchange
$\omega\leftrightarrow\omega'$ in the second term and recombine it
with the first one to obtain
\begin{multline}
\label{A2}
\frac{1}{D_{ \mathrm{ii}}}+\frac{1}{D_{ \mathrm {iii}}}
     +\frac{1}{D_{ \mathrm{viii}}}+\frac{1}{D_{ \mathrm {ix}}}
     +\frac{1}{D_{ \mathrm {iv}}}+\frac{1}{D_{ \mathrm{x}}}
\\
\to\frac{2}{\omega_A^n\!+\!\omega_B^m}
     \biggl(\frac{1}{\omega_A^n\!+\!\omega} +
\frac{1}{\omega_B^m\!+\!\omega} \biggr)
     \biggl(\frac{1}{\omega\!+\!\omega'}
      - \frac{1}{\omega\!-\!\omega'}
\biggr),
\end{multline}
where the symbol $\to$ denotes equality under the double frequency
integral. Similarly we have
\begin{multline}
\label{A3}
\frac{1}{D_{ \mathrm {i}}}+\frac{1}{D_{ \mathrm v}}
     +\frac{1}{D_{\mathrm {vi}}}
\\
    = \frac{1}{(\omega_A^n\!+\!\omega')
     (\omega_B^m\!+\!\omega')}
     \biggl(\frac{1}{\omega\!+\!\omega'} +
 \frac{1}{\omega\!-\!\omega'}
 \biggr)
\\
   - \frac{1}{(\omega_B^m\!+\!\omega')
 (\omega_A^n\!+\!\omega)(\omega\!-\!\omega')}\,,
\end{multline}
\begin{multline}
\label{A4}
\frac{1}{D_{ \mathrm{vii}}}\!+\!\frac{1}{D_{ \mathrm{xi}}}
     +\frac{1}{D_{\mathrm {xii}}}
\\
  =\frac{1}{(\omega_A^n\!\!+\omega')
     (\omega_B^m\!+\!\omega')}
     \biggl(\frac{1}{\omega\!+\!\omega'}
      + \frac{1}{\omega\!-\!\omega'}
 \biggr)
\\
   -\frac{1}{(\omega_A^n+\omega')(\omega_B^m+
\omega)(\omega-\omega')}\,.
\end{multline}
The second terms in Eqs.~(\ref{A3}) and (\ref{A4}) cancel each
other after an interchange of $\omega\leftrightarrow\omega'$ to yield
\begin{multline}
\label{A5}
    \frac{1}{D_{ \mathrm {i}}}+\frac{1}{D_{\mathrm v}}
     +\frac{1}{D_{ \mathrm{vi}}}
     +\frac{1}{D_{\mathrm {vii}}}+\frac{1}{D_{\mathrm {xi}}}
     +\frac{1}{D_{\mathrm{xii}}}
\\
    \to
     \frac{2}{(\omega_A^n\!+\!\omega)(\omega_B^m
     +\omega)}
     \biggl(\frac{1}{\omega\!+\!\omega'}
      - \frac{1}{\omega\!-\!\omega'}\biggr).
\end{multline}
Summation of Eqs. (\ref{A2}) and (\ref{A5})  immediately leads to
Eq.~(\ref{E21}).


\section{Scattering Green tensor for the planar multilayer system}
\label{mulg}

The scattering Green tensor for a planar multilayer system
can be given in the form \cite{chew}
\begin{equation}
\label{mul-a}
\bm{G}^{(1)}(\mathbf{r},\mathbf{r}',iu) =
\int\mathrm{d}^2q\,
e^{i\mathbf{q}\cdot(\mathbf{r}-\mathbf{r}')}
\bm{G}^{(1)}(\mathbf{q},z,z',iu)
\end{equation}
($\mathbf{q}\perp\mathbf{e}_z$), where
\begin{equation}
\label{mul-b}
\bm{G}^{(1)}(\mathbf{q},z,z',iu) = \frac{1}{8\pi^2b_N}
 \sum_{\sigma=s,p}\mathbf{e}_\sigma^+\mathbf{e}_\sigma^- r^\sigma_{N}
 e^{-b_N(z+z')},
 \end{equation}
with
\begin{align}
\label{es}
&\mathbf{e}_s^\pm=\sin{\phi}\,{\mathbf
e}_x-\cos\phi\,{\mathbf e}_y,
\\
\label{ep}
&\mathbf{e}_p^\pm=\mp\frac{b_N}{k_N}(\cos\phi\,{\mathbf e}_x+
\sin\phi\,{\mathbf e}_y)-\frac{iq}{k_N}\,{\mathbf e}_z
\end{align}
($\mathbf{e}_q$ $\!=$ $\!\cos\phi\,\mathbf{e}_x$
$\!+$ $\!\sin\phi\,\mathbf{e}_y$ $\!=$ $\!\mathbf{q}/q$, $q$ $\!=$
$\!|\mathbf{q}|$) denoting the polarization vectors for $s$- and
$p$-polarized waves propagating in the positive($+$)/negative($-$)
$z$-direction. Further, $b_N$ and $k_N$, respectively, are defined
according to Eqs.~(\ref{bl}) and (\ref{kl}), and the generalized
reflection coefficients are given in Eq.~(\ref{rsp}).
Equations~(\ref{es}) and (\ref{ep}) imply that
\begin{equation}
\mathbf e_{s}^+\mathbf e_{s}^{-}=
\left(
\begin{array}{lll}
\sin^2\phi & -\sin\phi\cos\phi & 0 \\
-\sin\phi\cos\phi & \cos^2\phi & 0 \\
0 & 0 & 0
\end{array}
\right),
\end{equation}
\begin{eqnarray}
&&\mathbf e_{p}^+\mathbf e_{p}^{-}=\nonumber\\[1ex]
&&\!\left(
\begin{array}{lll}
-\frac{b_N^2}{k_N^2}\cos^2\phi & -\frac{b_N^2}{k_N^2}
\sin\phi\cos\phi & \frac{ib_Nq}{k_N^2}\cos\phi \\
-\frac{b_N^2}{k_N^2}\sin\phi\cos\phi &-\frac{b_N^2}{k_N^2}
\sin^2\phi &\frac{ib_Nq}{k_N^2}\sin\phi \\
-\frac{ib_Nq}{k_N^2}\cos\phi & -\frac{ib_Nq}{k_N^2}\sin\phi  &
-\frac{q^2}{k_N^2}
\end{array}
\right).\nonumber\\
\end{eqnarray}
Substituting these results into Eqs.~(\ref{mul-a}) and (\ref{mul-b}),
performing the $\phi$-integrals by means of~\cite{abra}
\begin{equation}
\int_0^{2\pi}{\mathrm d}x\,e^{iz\cos{x}}\cos(\nu x)=2\pi
i^{\nu} J_{\nu}(x),
\end{equation}
and using the relation
\begin{equation}
\frac{J_1(x)}{x}=\frac{J_0(x)-J_2(x)}{2}\,,
\end{equation}
we arrive at the Eqs.~(\ref{62})--(\ref{64}).

In the particular case of a perfectly reflecting plate in the retarded
limit, it is convenient to replace the integration variable $q$
in Eqs.~(\ref{62})--(\ref{64}) in favour of \mbox{$v$ $\!=$
$\!b_1c/u$}, i.e., $q$ $\!=$ $\!\sqrt{v^2-1}u/c$ [see Eq.~(\ref{bl})],
and hence
\begin{equation}
\label{C8}
\int_0^\infty{\mathrm d}q\,\frac{q}{b_1}\cdots
\ \mapsto\ \int_1^\infty
 {\mathrm d}v\,\frac{u}{c}\cdots\,.
\end{equation}
For $X$ $\!\ll$ $\!Z_+$, the exponential terms effectively limits the
integrals in Eqs.~(\ref{62})--(\ref{64}) to the region where
\mbox{$qX$ $\!\ll$ $\!1$}, hence we can approximate \mbox{$J_\nu(qX)$
by $J_\nu(0)$ $\!=$ $\!\delta_{\nu 0}$}, such that the nonzero
scattering-Green tensor components read
\begin{align}
\label{rx}
&G^{(1)}_{xx}(\mathbf r_{A},\mathbf r_{B},iu)
 =G^{(1)}_{yy}(\mathbf r_{A},\mathbf r_{B},iu)\nonumber\\
&\quad =\frac{1}{8\pi Z_+}\bigg[r_s-\bigg(1+2\frac{c}{Z_+u}+
 2\frac{c^2}{Z_+^2u^2}\bigg)r_p\bigg]e^{-Z_+u/c},\\
\label{rz}
&G^{(1)}_{zz}(\mathbf r_{A},\mathbf r_{B},iu)\nonumber\\
&\quad =-\frac{1}{2\pi Z_+}\bigg(\frac{c}{Z_+u}+
 \frac{c^2}{Z_+^2u^2}\bigg)r_p\,e^{-Z_+u/c},
\end{align}
leading to Eqs.~(\ref{gl79}) and (\ref{gl80}), recall
Eq.~(\ref{alpha0}).

In the nonretarded limit it can be shown that the main contribution to
the frequency integrals comes from the region where $u/(cb_1)$ $\!\ll$
$\!1 $ (cf. Ref.~\cite{Thomas}). In this region we have
\begin{equation}
\label{9}
q=b_1\sqrt{1-\frac{u^2}{b_1^2c^2}}\simeq b_1\equiv b.
\end{equation}
By changing the integration variable $q$ according to
\begin{equation}
\label{change}
 \int_0^\infty{\rm d}q\,\frac{q}{b_1}\,
 \ldots
\ \mapsto\
\int_{u/c}^\infty\mathrm{d}b\,
\ldots
\end{equation}
and setting the lower limit of integration to zero, from
Eqs.~(\ref{62})--(\ref{64}) we find, after some algebra, the nonzero
elements of the scattering Green tensor to be approximately given by
\begin{align}
\label{xx}
& G^{(1)}_{xx}(\mathbf r_{A},\mathbf r_{B},iu)=
\frac{c^2}{4\pi
u^2}\frac{2X^2-Z_+^2}{l_+^5}\,r_p,\quad\\
\label{yy}
& G^{(1)}_{yy}(\mathbf r_{A},\mathbf r_{B},iu)=-
\frac{c^2}{4\pi
u^2}\frac{1}{l_+^3}\,r_p,\quad\\
\label{xz}
& G^{(1)}_{xz(zx)}
(\mathbf r_{A},\mathbf
r_{B},iu)=\MP
\frac{c^2}{4\pi u^2 }\frac{3XZ_+}{l_+^5}\,r_p,\\
\label{zz}
& G^{(1)}_{zz}(\mathbf r_{A},\mathbf r_{B},iu)=
\frac{c^2}{4\pi u^2}\frac{X^2-2Z_+^2}{l_+^5}\,r_p,
\end{align}
with $l_+$ $\!=$ $\!\sqrt{X^2+Z_+^2}$, leading to Eqs.~(\ref{e1}) and
(\ref{e2}).

For a semi-infinite magnetodielectric half space in the nonretarded
limit, we apply a similar procedure as below Eq.~(\ref{rz}) and expand
the reflection coefficients given by Eq.~(\ref{eq95}) in terms of
$u/(bc)$,
\begin{align}
\label{rs-lim}
&r_s\simeq\frac{\mu(iu)-1}{\mu(iu)+1}-\frac{\mu(iu)[
\varepsilon(iu)\mu(iu)-1]}{[\mu(iu)+1]^2}\frac{u^2}{b^2c^2}\,,
\\
\label{rp-lim}
&r_p\simeq\frac{\varepsilon(iu)-1}{\varepsilon(iu)+1}-\frac{
\varepsilon(iu)[\varepsilon(iu)\mu(iu)-1]}{[\varepsilon(iu)+1]^2}
\frac{u^2}{b^2c^2}\,.
\end{align}
Substituting (\ref{rs-lim}) and (\ref{rp-lim}) into
Eqs.~(\ref{62})--(\ref{64}) and keeping only the leading-order terms
of $u/bc$, in the case of the purely dielectric half space we can
ignore $r_s$ and the second term in the r.h.s.\ of Eq.~(\ref{rp-lim}),
so the relevant elements of the scattering Green tensor are
approximately 
\begin{align}
\label{Axx}
 G^{(1)}_{xx}(\mathbf r_{A},\mathbf r_{B},iu)
 =\frac{2X^2-Z_+^2}{4\pi
 l_+^5}\,\frac{\varepsilon(iu)-1}{\varepsilon(iu)+1}\frac{c^2}{u^2}\,,
\end{align}
\begin{equation}
\label{Byy}
 G^{(1)}_{yy}(\mathbf r_{A},\mathbf r_{B},iu)
 =-\frac{1}{4\pi l_+^3}
\frac{\varepsilon(iu)-1}{\varepsilon(iu)+1}\frac{c^2}{u^2}\,,
\end{equation}
\begin{align}
\label{Axz}
G^{(1)}_{xz}(\mathbf r_{A},\mathbf r_{B},iu)
& 
=- G^{(1)}_{zx}(\mathbf r_{A},\mathbf r_{B},iu)\nonumber\\
&
=-\frac{3XZ_+}{4\pi l_+^5}
\frac{\varepsilon(iu)-1}{\varepsilon(iu)+1}\frac{c^2}{u^2}\,,
\end{align}
\begin{equation}
\label{Bzz}
 G^{(1)}_{zz}(\mathbf r_{A},\mathbf r_{B},iu)
 =\frac{X^2-2Z_+^2}{4\pi l_+^5}
\frac{\varepsilon(iu)-1}{\varepsilon(iu)+1}\frac{c^2}{u^2}\,.
\end{equation}
For a purely magnetic half space, the first term on the r.h.s.\ of
Eq.~(\ref{rp-lim}) vanishes, so the leading order of $u/bc$ is due to
the second term as well as the first term on the r.h.s.\ of
Eq.~(\ref{rs-lim}), so the nonzero elements of the scattering Green
tensor can be approximated by
\begin{align}
\label{Axx2}
 G^{(1)}_{xx}(\mathbf r_{A},\mathbf r_{B},iu)
 =&\;\frac{ l_+-Z_+}{4 \pi X^2}
\frac{\mu(iu)-1}{\mu(iu)+1}
\nonumber\\&
+\frac{Z_+l_+-Z_+^2}{16\pi X^2l_+}
[\mu(iu)-1],
\end{align}
\begin{align}
\label{Ayy}
 G^{(1)}_{yy}(\mathbf r_{A},\mathbf r_{B},iu)=&\;
 \frac{l_+-Z_+}{16 \pi  X^2}
 [\mu(iu)-1]
\nonumber\\&
+\frac{Z_+l_+-Z_+^2}{4 \pi X^2l_+}
\frac{\mu(iu)-1}{\mu(iu)+1}\,,
\end{align}

\begin{align}
\label{Axz2}
G^{(1)}_{xz}(\mathbf r_{A},\mathbf r_{B},iu)
&
=- G^{(1)}_{zx}(\mathbf
 r_{A},\mathbf r_{B},iu)\nonumber\\
&
=\frac{l_+-Z_+}{16 \pi Xl_+}
 [\mu(iu)-1],
\end{align}
\begin{equation}
\label{Azz}
 G^{(1)}_{zz}(\mathbf r_{A},\mathbf r_{B},iu)=\frac{1}{16 \pi l_+}
[\mu(iu)-1].
\end{equation}


\section{Explicit forms of $\bm{A_{n\pm}}$ and $\bm{B_n}$ in
Eqs.~(\ref{A}) and (\ref{B})}
\label{an}

The integrals in Eqs.~(\ref{A}) and (\ref{B}) can be performed to
obtain the following explicit expressions:
\begin{align}
&A_{3+}=\frac{6a}{\bigl(a^2+\beta^2\bigr)^\frac{5}{2}}\,,
\\
&A_{3-}=\frac{6\bigl(a^3-4a \beta^2\bigr)}
 {\bigl(a^2+\beta^2\bigr)^\frac{7}{2}}\,,\\
&A_{4+}=
\frac{6\bigl(4a^2-\beta^2\bigr)}
{\bigl(a^2+\beta^2\bigr)^\frac{7}{2}}\,,\\
&A_{4-}= \frac{6\bigl(4a^4-27a^2\beta^2+4\beta^4\bigr)}
 {\bigl(a^2+\beta^2\bigr)^\frac{9}{2}}\,,\\
&A_{5+}=\frac{30\bigl(4a^3-3a\beta^2\bigr)}
{\bigl(a^2+\beta^2\bigr)^\frac{9}{2}}\,,\\
& A_{5-}=
 \frac{30\bigl(4a^5-41a^3\beta^2+18a\beta^4\bigr)}
 {\bigl(a^2+\beta^2\bigr)^\frac{11}{2}}\,,\\
&B_{3}=\frac{3a\bigl(2a^2-3\beta^2\bigr)}
{\bigl(a^2+\beta^2\bigr)^\frac{7}{2}}\,,\\
&B_{4}=\frac{3\bigl(8a^4-24a^2\beta^2+3\beta^4\bigr)}
{\bigl(a^2+\beta^2\bigr)^\frac{9}{2}}\,,\\
&B_{5}=\frac{15a\bigl(8a^4-40a^2\beta^2+15\beta^4\bigr)}
{\bigl(a^2+\beta^2\bigr)^\frac{11}{2}}\,.
\end{align}

\end{appendix}


\end{document}